\documentclass[a4paper,11pt]{article}
\usepackage{jcappub} 
\usepackage{lineno}
\usepackage{graphicx}
\usepackage{amssymb}
\usepackage{bm}
\usepackage{color}
\usepackage{multirow}
\usepackage{subfigure}
\usepackage{array}
\usepackage{cases}
\newcommand{\f}{\frac}
\newcommand{\lt}{\left}

\newcommand{\n}{\nonumber}

\newcommand{\rt}{\right}
\newcommand{\dd}{{\rm d}}

\newcommand{\dt}{\delta}
\newcommand{\ep}{\epsilon}

\newcommand{\sg}{\sigma}

\newcommand{\pb}{{\rm PBH}}

\title{\boldmath Primordial black hole clustering from spectator fields for interpreting the JWST observations}

\author[a,b]{Bing-Yu Su,}
\author[c,*]{Nan Li,\note{Corresponding author.}}
\author[a,b,*]{and Lei Feng}
\affiliation[a]{Key Laboratory of Dark Matter and Space Astronomy, Purple Mountain Observatory, Chinese Academy of Sciences, Nanjing 210023, China}
\affiliation[b]{School of Astronomy and Space Science, University of Science and Technology of China, Hefei, Anhui 230026, China}
\affiliation[c]{Department of Physics, College of Sciences, Northeastern University, Shenyang 110819, China}

\emailAdd{linan@mail.neu.edu.cn, fenglei@pmo.ac.cn}

\abstract{The observations by the James Webb Space Telescope (JWST) have revealed unexpectedly massive galaxy candidates at high redshifts, posing a significant challenge to the $\Lambda$CDM model. In this work, we investigate whether primordial black holes (PBHs) with spatial clustering, generated by a light spectator field during inflation, can accelerate early structure formation. We adopt the galaxy candidates with inferred stellar mass $10^9\,M_\odot\leq M_*^{\rm obs}\leq10^{11}\,M_\odot$ at redshift $7 \leq z \leq 10$ reported by the CEERS program as a benchmark. Two different mechanisms are considered, through which PBH clustering can influence structure formation: the PBH-induced isocurvature perturbations that enhance the matter power spectrum on linear scales, and the localized seed formation and accretion by compact PBH clusters on nonlinear scales. We find that, when adopting the cosmic microwave background (CMB) isocurvature constraint $\beta_{\rm iso}<0.035$ at the benchmark pivot scale $k_*=0.002\,{\rm Mpc}^{-1}$, PBH clustering can produce a cumulative stellar mass density consistent with the JWST observations while satisfying the relevant isocurvature constraint. However, the allowed enhancement of structure formation is strongly suppressed when the constraint at $k_*=0.1\,{\rm Mpc}^{-1}$ is imposed, indicating a significant dependence on the choice of the pivot scale. In contrast, the localized seed effect of compact PBH clusters is strongly constrained by the CMB isocurvature bounds, while isolated supermassive PBHs produce stellar mass densities far below those inferred from the JWST observations. Our results show that PBH clustering induced by a spectator field can substantially accelerate early structure formation, but whether it can fully account for the JWST-inferred stellar mass density depends sensitively on the pivot scale adopted for the CMB isocurvature constraint. }

\begin{document}
\maketitle
\flushbottom

\section{Introduction} \label{sec:intro}

The remarkable capability of the James Webb Space Telescope (JWST) has opened a new window into the Universe at the end of the cosmic dark ages. The early JWST observations revealed an unexpectedly large population of bright and massive galaxy candidates at high redshifts $z$~\cite{2022ApJ...940L..14N, adams2023discovery, Yan:2022sxd, 2023MNRAS.519.1201A, 2023MNRAS.518.6011D, Labbe:2022ahb, 2022ApJ...940L..55F, 2023ApJS..265....5H, Robertson:2022gdk, Robertson:2023nyv, Perez-Gonzalez:2023wta}. In particular, the Cosmic Evolution Early Release Science (CEERS) program identified a population of candidate galaxies with inferred stellar masses spanning approximately $10^9$--$10^{11}\,M_\odot$ ($M_\odot =1.99\times 10^{33}$ g is the solar mass) over $7\leq z\leq10$~\cite{Labbe:2022ahb}. The subsequent JWST spectroscopic observations have confirmed the existence of such galaxies at even earlier cosmic times~\cite{2023NatAs...7..611R, 2023NatAs...7..622C, 2023ApJ...951L..22A} with $z\sim14$~\cite{2024Natur.633..318C}, demonstrating that substantial galaxy populations were already in place within the first few hundred million years after the Big Bang. 

These observations have raised significant challenges for standard astrophysical models and prompted the discussions on the potential tension with the conventional $\Lambda$ cold dark matter ($\Lambda$CDM) picture of hierarchical structure formation~\cite{2013ApJ...770...57B, 2019MNRAS.488.3143B, 2022ApJ...938L..10I, Lovell:2022bhx, Forconi:2023izg}. Various observational and astrophysical scenarios have been proposed to alleviate this tension~\cite{Menci:2022wia, Gong:2022qjx, Biagetti:2022ode, Wang:2022jvx, Parashari:2023cui, Dayal:2023nwi, 2023MNRAS.526.2542C, Lopez-Corredoira:2024pgl}. On the observational side, uncertainties in photometric-redshift determination and spectral-energy-distribution modeling~\cite{2026arXiv260418683P}, including assumptions about dust attenuation~\cite{Ferrara:2022dqw, Ziparo:2022rir, 2025A&A...702A..33M} and star formation histories~\cite{2023A&A...679A..12M, 2025NatAs...9..458M}, can affect the inferred redshifts and stellar masses of high redshift candidates. Spectroscopic follow-up has indeed identified a fraction of lower-redshift interlopers among photometrically selected extreme high-redshift candidates~\cite{2023Natur.622..707A}, while simultaneously confirming a substantial population of genuinely high-redshift galaxies~\cite{2023NatAs...7..611R, 2023NatAs...7..622C, 2023ApJ...951L..22A, 2024Natur.633..318C}. Cosmic variance and the limited survey volume may also influence the inferred abundance of rare massive galaxies~\cite{2023MNRAS.526.2542C}. On the astrophysical side, enhanced star formation efficiencies~\cite{Boylan-Kolchin:2022kae, 2023MNRAS.519..843M}, dark energy parameterization~\cite{Wang:2023ros}, feedback-free starbursts~\cite{Dekel:2023ddd}, and alternative feedback mechanisms~\cite{2023ApJ...955L..35S} have been explored as possible mechanisms for accelerating galaxy growth in the early Universe. 

Nevertheless, whether these astrophysical effects alone are sufficient to account for the observations remains an open question. This motivates us to consider primordial black holes (PBHs), which can theoretically form through the gravitational collapse of sufficiently large overdensities in the early Universe~\cite{Zeldovich:1967lct, Hawking:1971ei, Kohri:2014lza}, as a complementary mechanism for accelerating the assembly of cosmic structures~\cite{Liu:2022bvr, Yuan:2023bvh, Huang:2024aog, Hooper:2023nnl, Dave:2026gsd}. PBHs can enhance structure formation through the perturbations in their spatial number density~\cite{Meszaros:1975ef}. These perturbations correspond to isocurvature (or entropy) perturbations, which initially arise without accompanying radiation density perturbations. After matter--radiation equality, they can source matter density perturbations and enhance the growth of cosmic structures through gravitational instability. PBHs can also promote structure formation through the distinct seed effect~\cite{1983ApJ...268....1C, Carr:1984ikr, Carr:2018rid}, whereby the gravitational attraction of sufficiently massive PBHs locally amplifies the accretion and collapse of surrounding matter. The two mechanisms operate in different regimes: the isocurvature mechanism enhances matter fluctuations on linear scales, whereas the seed effect provides a localized enhancement of structure formation around sufficiently massive PBH seeds. Therefore, both mechanisms can facilitate the growth of DM halos and, ultimately, the formation of galaxies.

However, interpreting the JWST observations within conventional PBH scenarios faces severe theoretical and observational constraints. In particular, explaining the excess of massive galaxies through either the Poisson or seed effect requires very massive PBHs with $M_{\rm PBH}\gtrsim 10^4\,M_\odot$~\cite{Liu:2022bvr, Hutsi:2022fzw}. In the simplest single field inflationary scenarios with approximately Gaussian primordial curvature perturbations~\cite{Kannike:2017bxn, Ballesteros:2017fsr, Gangopadhyay:2021kmf, Wang:2021kbh, Liu:2021qky}, the formation of PBHs generally requires a substantial enhancement of the primordial power spectrum on scales much smaller than those probed by the anisotropies of the cosmic microwave background (CMB)~\cite{Planck:2018vyg, Planck:2018jri}. For sufficiently massive PBHs, the corresponding enhancement can extend to scales relevant for the CMB spectral distortions. The subsequent dissipation of these enhanced curvature perturbations into the primordial plasma prior to recombination may generate the so-called $\mu$-type spectral distortions in the CMB, thereby imposing stringent constraints on the amplitude of the primordial power spectrum~\cite{Carr:2020gox}. Additional constraints on the PBH abundance obtained from Lyman-$\alpha$ forest~\cite{Murgia:2019duy}, dynamical friction~\cite{1999ApJ...516..195C}, and X-ray binaries~\cite{Inoue:2017csr} together severely restrict the parameter space of conventional PBH scenarios, making such PBH explanation of the JWST observations rather disfavored~\cite{1983ApJ...268....1C}. 

In this work, we instead consider the PBHs generated through a light spectator field during inflation, which provides a way to circumvent the above constraints, while substantially reducing the PBH mass and abundance required to account for the observed early structure formation~\cite{Nakama:2016kfq}. In the spectator field framework, the observable Universe consists of a dominant population of patches following the standard inflationary history driven by the slow-roll inflaton field and a subdominant population of patches that undergo additional expansion driven by the spectator field. The latter can develop sufficiently large primordial perturbations to collapse into PBHs, while the perturbations in the vast majority of regions remain close to the nearly scale-invariant spectrum consistent with the CMB observations. 

Our mechanism possesses two advantages. On the one hand, PBHs are not spatially distributed as a Poisson process. The long-wavelength fluctuations of the spectator field modulate the local probability of PBH formation, so that the regions with similar values of the spectator field are more likely to have similar PBH abundance. As a result, PBHs tend to exist preferentially in the same spatial regions rather than being independently distributed, giving rise to strong spatial clustering~\cite{Shinohara:2021psq}. A variety of mechanisms from PBH clustering have also been investigated~\cite{Belotsky:2018wph, Colazo:2024jmz, Khlopov:2004sc, Barcelo-Sastre:2026nbs, Zhang:2026uum}. Such clustering corresponds to an additional PBH isocurvature perturbation and can substantially enhance the matter perturbations on small scales~\cite{Desjacques:2018wuu, Matsubara:2019qzv, Kawasaki:2021zir, Huang:2024aog}, requiring lower PBH mass and overall abundance compared to the Poisson effect arising from unclustered PBHs. On the other hand, because the large perturbations required for PBH formation are confined to a small fraction of the Universe, rather than appearing as a global enhancement of the primordial power spectrum, our model can produce even supermassive PBHs without violating the COBE/FIRAS constraints on the $\mu$-distortions~\cite{Nakama:2016kfq}. As a result, the gravitational growth of these clustered PBHs can accelerate the formation of DM halos and their host galaxies. This provides a natural possibility for explaining the early emergence of massive galaxies via the PBHs with lower masses. 

This paper is organized as follows. In Sec.~\ref{sec:spec}, we introduce the spectator field model and derive the spatial two-point correlation function and the clustering properties of PBHs. In Sec.~\ref{sec:iso}, we calculate the primordial isocurvature power spectrum induced by clustered PBHs and evaluate the constraints from the Planck CMB constraints. The enhancement of early halo formation induced by PBH clustering through the isocurvature perturbations and the seed effect is discussed in Sec.~\ref{sec:clustering}, where we also examine whether such mechanisms can account for the JWST observations. Finally, we summarize our main conclusions in Sec.~\ref{sec:con}. Throughout this paper, we adopt the system of natural units with $c=\hbar=k_{\rm B}=1$.

\section{PBHs from a spectator field} \label{sec:spec}

Within the spectator field framework, the spatial fluctuations of light spectator field $\chi$ generated during inflation modulate the local probability of PBH formation. The regions in which the spectator field $\chi$ exceeds a critical value $\chi_{\rm c}$ can enter the PBH forming regime, while the regions with $\chi<\chi_{\rm c}$ produce no PBHs. Since the fluctuations of the spectator field are spatially correlated, the probability of PBH formation at one location is correlated with another, leading to a non-Poisson spatial distribution of PBHs. Therefore, we need to characterize the PBH clustering by the PBH two-point correlation function $\xi_{\rm PBH}(r)$. This function measures the excess probability of finding a PBH pair separated by a distance $r$ relative to a spatially uncorrelated distribution. In particular, $\xi_{\rm PBH}(r)>0$ indicates an enhanced probability of finding PBH pairs and thus corresponds to positive spatial clustering. 

The PBH two-point correlation function $\xi_{\rm PBH}(r)$ is determined by the one- and two-point formation probabilities of the spectator field. We define $\chi_{\rm ini}$ as the value of the spectator field $\chi$ at the time when the observable Universe exited the Hubble horizon. PBHs then form in regions where $\chi({\bf x})-\chi_{\rm ini}>\chi_{\rm c}$ evaluated at the end of the inflation. Then, the probability of PBH formation at the origin point ${\bf x}={\bf 0}$ is~\cite{Shinohara:2021psq}
\begin{align}
P_1({\bf 0}) =\int[D\chi]\,P[\chi] \int_{\chi_{\rm c}}^\infty\dt_{\rm D}(\chi({\bf 0})-\alpha_1)\,\dd\alpha_1, \label{P1}
\end{align}
where $[D\chi] =\prod_{i=1}\dd\chi({\bf x}_i)$ is the integral measure, $P[\chi]$ is the probability functional associated with the field configuration $\chi({\bf x})$, $\chi_{\rm c}$ is the threshold of the spectator field for PBH formation, and $\dt_{\rm D}$ is the Dirac $\delta$ function, which can be reformulated as 
\begin{align}
\dt_{\rm D}(\chi({\bf x})-\alpha)=\int_{-\infty}^\infty e^{iq(\chi({\bf x})-\alpha)}\,\f{\dd q}{2\pi}. \label{dtD}
\end{align} 
Similarly, the joint probability that PBHs are formed simultaneously at two spatial points ${\bf 0}$ and ${\bf x}$ is given by
\begin{align}
P_2({\bf 0}; {\bf x}) =\int[D\chi]\,P[\chi]\int_{\chi_{\rm c}}^\infty\dt_{\rm D}(\chi({\bf 0})-\alpha_1)\,\dd\alpha_1 \int_{\chi_{\rm c}}^\infty\dt_{\rm D}(\chi({\bf x})-\alpha_2)\,\dd\alpha_2. \label{P2}
\end{align}
Following Ref.~\cite{Shinohara:2021psq}, after some algebra, we can obtain the following expressions,
\begin{align}
P_1({\bf 0})&=\f{1}{2}{\rm erfc}\lt(\f{\nu}{\sqrt{2}}\rt), \\ 
P_2({\bf 0};{\bf x})&=\f{1}{2\sqrt{2\pi}}\int_\nu^\infty e^{-\tilde\nu^2/2}{\rm erfc}\lt[\f{\nu-\ep \tilde\nu}{\sqrt{2(1-\ep^2)}}\rt]\,\dd\tilde\nu, 
\end{align}
where ${\rm erfc}(x)$ is the complementary error function, and the parameters $\nu$ and $\ep$ are defined as
\begin{align}
\nu=\f{\chi_{\rm c}}{\sqrt{\xi_\chi({\bf 0},{\bf 0})}}, \quad \ep=\f{\xi_\chi({\bf 0},{\bf x})}{\xi_\chi({\bf 0},{\bf 0})}, \label{nuep}
\end{align}
with $\xi_\chi$ being the two-point correlation function of the spectator field $\chi$, and $|\ep|\leq 1$ being the normalized two-point correlation coefficient of $\chi$. 

In the following, we denote $r=|{\bf x}|$, and then $\xi_\chi(r)$ should be evaluated at the time of the end of inflation $t_{\rm end}$ as
\begin{align}
\xi_\chi(r)=\langle(\chi(t_{\rm end},{\bf 0})-\chi_{\rm ini})(\chi(t_{\rm end},{\bf x})-\chi_{\rm ini})\rangle=\int_{k_{\rm min}}^{k_{\rm max}}{\cal P}_\chi(k)\f{\sin(kr)}{kr}\,\f{\dd k}{k}, \label{xichir}
\end{align}
where $\chi_{\rm ini}$ is the horizon exit value, and the dimensionless power spectrum ${\cal P}_\chi(k)$ of the spectator field $\chi$ with mass $m_\chi$ reads
\begin{align}
{\cal P}_\chi(k)=\lt(\f{H_{\rm inf}}{2\pi}\rt)^2\lt(\f{k}{k_{\rm end}}\rt)^{c_{\rm inf}}, \label{Pchik}
\end{align}
where $H_{\rm inf}$ is the Hubble rate during inflation (assumed to be constant), $k_{\rm end}\simeq10^{19}\,{\rm Mpc}^{-1}$ is the wave number at the end of inflation~\cite{Shinohara:2021psq}, and 
\begin{align}
c_{\rm inf}=\f{2m_\chi^2}{3H_{\rm inf}^2}
\end{align}
is a dimensionless parameter, respectively. As the spectator $\chi$ is assumed to be light, with $m_\chi\ll H_{\rm inf}$, we have $c_{\rm inf}\ll1$. 
Since we study the correlation of $\chi$ inside the observable Universe, the minimum wave number $k_{\rm min}$ is taken as the present Hubble scale as $k_{\rm min} = H_0$ in the following calculations, with $H_0=67.36~{\rm km}~{\rm s}^{-1}~{\rm Mpc}^{-1}$ being the Hubble constant~\cite{Planck:2018vyg}. We integrate over the modes from the present Hubble scale with $k_{\min}=H_0$ up to a maximum wavenumber $k_{\max}$ corresponding to the comoving scale associated with PBHs of mass $M_{\rm PBH}$~\cite{Mishra:2019pzq, Carr:2020gox},
\begin{align}
k_{\rm max}\simeq 2\times10^6 \lt(\f{\gamma}{0.2}\rt)^{1/2}\lt(\f{g_{*,{\rm form}}}{10.75}\rt)^{-1/12}\lt(\f{M_\pb}{M_\odot}\rt)^{-1/2}\,{\rm Mpc}^{-1}, \label{kmax}
\end{align}
where $\gamma\simeq 0.2$ is the efficiency of collapse, and $g_{*,{\rm form}}$ is the relativistic effective degree of freedom for energy density, which evolves with the cosmic temperature $T$ and its specific values can be obtained by interpolating the data in Ref.~\cite{Saikawa:2018rcs}. Moreover, the dependence of $M_\pb$ on $g_{*,{\rm form}}$ and $T$ reads~\cite{Carr:2020gox}
\begin{align}
{M_{\rm PBH}}=2.03\times10^5\lt(\f{\gamma}{0.2}\rt)\lt(\f{g_{*,{\rm form}}}{10.75}\rt)^{-1/2}\lt(\f{T}{{\rm MeV}}\rt)^{-2}~{M_\odot}. 
\end{align}
Now, substituting Eq.~\eqref{Pchik} into Eq.~\eqref{xichir}, we obtain the self-correlation function of the spectator field $\chi$ as
\begin{align}
\xi_\chi(0)=\lt(\f{H_{\rm inf}}{2\pi}\rt)^2\f{1}{c_{\rm inf}}\lt[\lt(\f{k_{\rm max}}{k_{\rm end}}\rt)^{c_{\rm inf}}-\lt(\f{k_{\rm min}}{k_{\rm end}}\rt)^{c_{\rm inf}}\rt].
\end{align}
Then, from the definition of $\ep$ in Eq.~\eqref{nuep}, we have~\cite{Shinohara:2021psq}
\begin{align}
\ep(r) =c_{\rm inf}\lt[\lt(\f{k_{\rm max}}{k_{\rm end}}\rt)^{c_{\rm inf}}-\lt(\f{k_{\rm min}}{k_{\rm end}}\rt)^{c_{\rm inf}}\rt]^{-1}\int_{k_{\rm min}}^{k_{\rm max}}\lt(\f{k}{k_{\rm end}}\rt)^{c_{\rm inf}}\f{\sin(kr)}{kr}\f{\dd k}{k}. 
\end{align} 

The last ingredient is to fix the parameter $\nu$. We determine $\nu$ by requiring the PBH abundance $f_{\rm PBH}$, defined as the present day fraction of DM in PBHs, to reproduce the value needed in our structure formation scenario. 
In this work, we assume the PBH mass function to be monochromatic (i.e., all PBHs possess the same mass $M_\pb$), and then $f_\pb$ can be expressed as~\cite{Sasaki:2018dmp}
\begin{align}
f_\pb&=2.7\times10^8\lt(\f{\gamma}{0.2}\rt)^{1/2}\lt(\f{g_{*,{\rm form}}}{10.75}\rt)^{-1/4}\lt(\f{M_\pb}{M_\odot}\rt)^{-1/2}P_1. \label{fPBH}
\end{align}
Furthermore, $f_\pb$ can be related to the average comoving number density $\bar{n}_{\rm PBH}$ as
\begin{align}
\bar{n}_{\rm PBH}=\f{3H_0^2\Omega_{\rm DM}}{8\pi G} 
\f{f_{\rm PBH}}{M_{\rm PBH}} \simeq3.25\times10^{10}f_{\rm PBH} \left(\f{M_{\rm PBH}}{M_\odot}\right)^{-1}\,{\rm Mpc}^{-3}, \label{nPBH}
\end{align}
where $\Omega_{\rm DM}=0.266$ is the present day DM density parameter~\cite{Planck:2018vyg}. For a given PBH mass, the required PBH abundance $f_\pb$ can be inferred from $\bar{n}_{\rm PBH}$. Then, We can determine $\nu$ by requiring the corresponding comoving PBH number density to reproduce this target abundance through Eqs.~\eqref{fPBH} and \eqref{nPBH}.

\begin{figure}[htb]
\centering \includegraphics[width=0.48\textwidth]{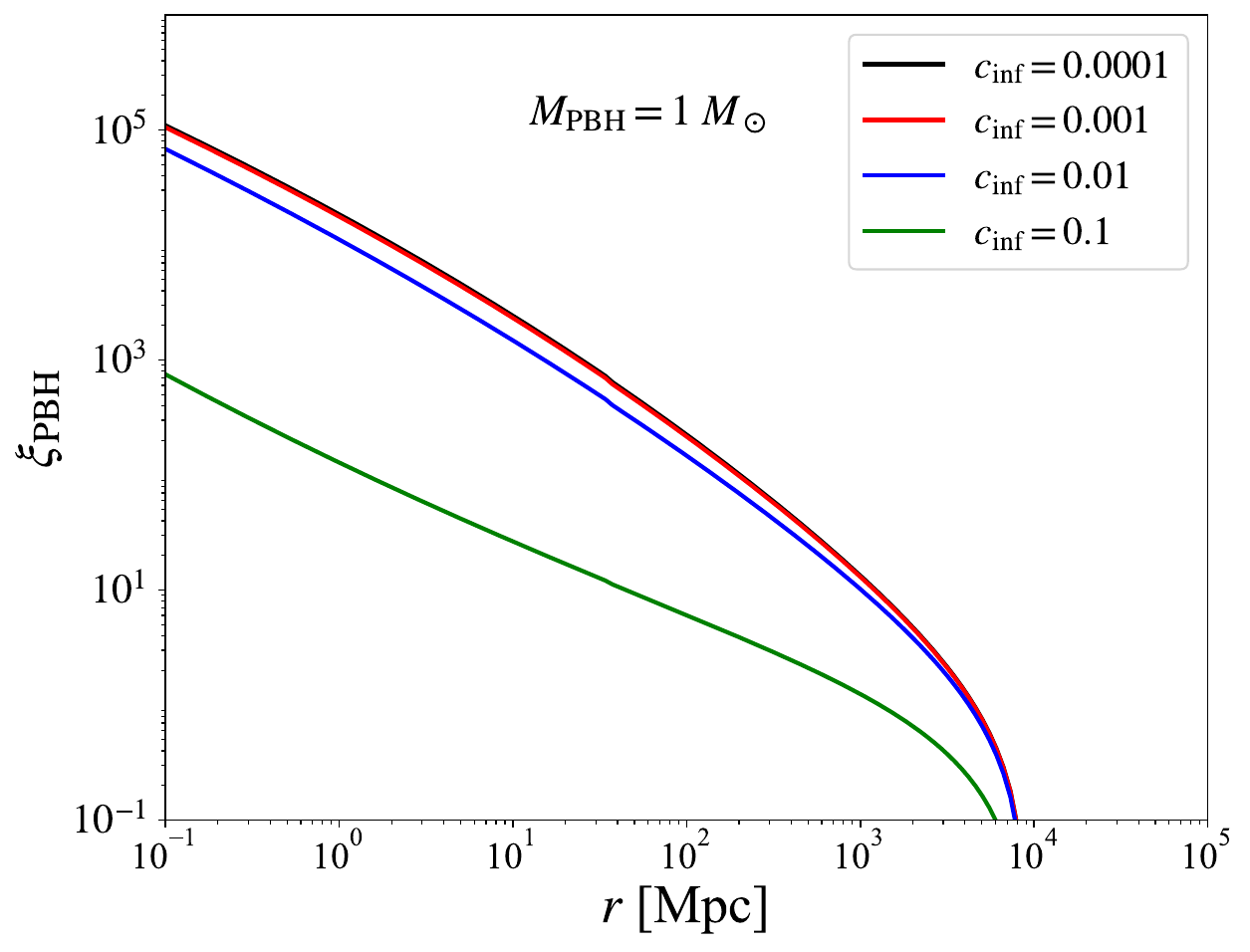} \quad
\includegraphics[width=0.48\textwidth]{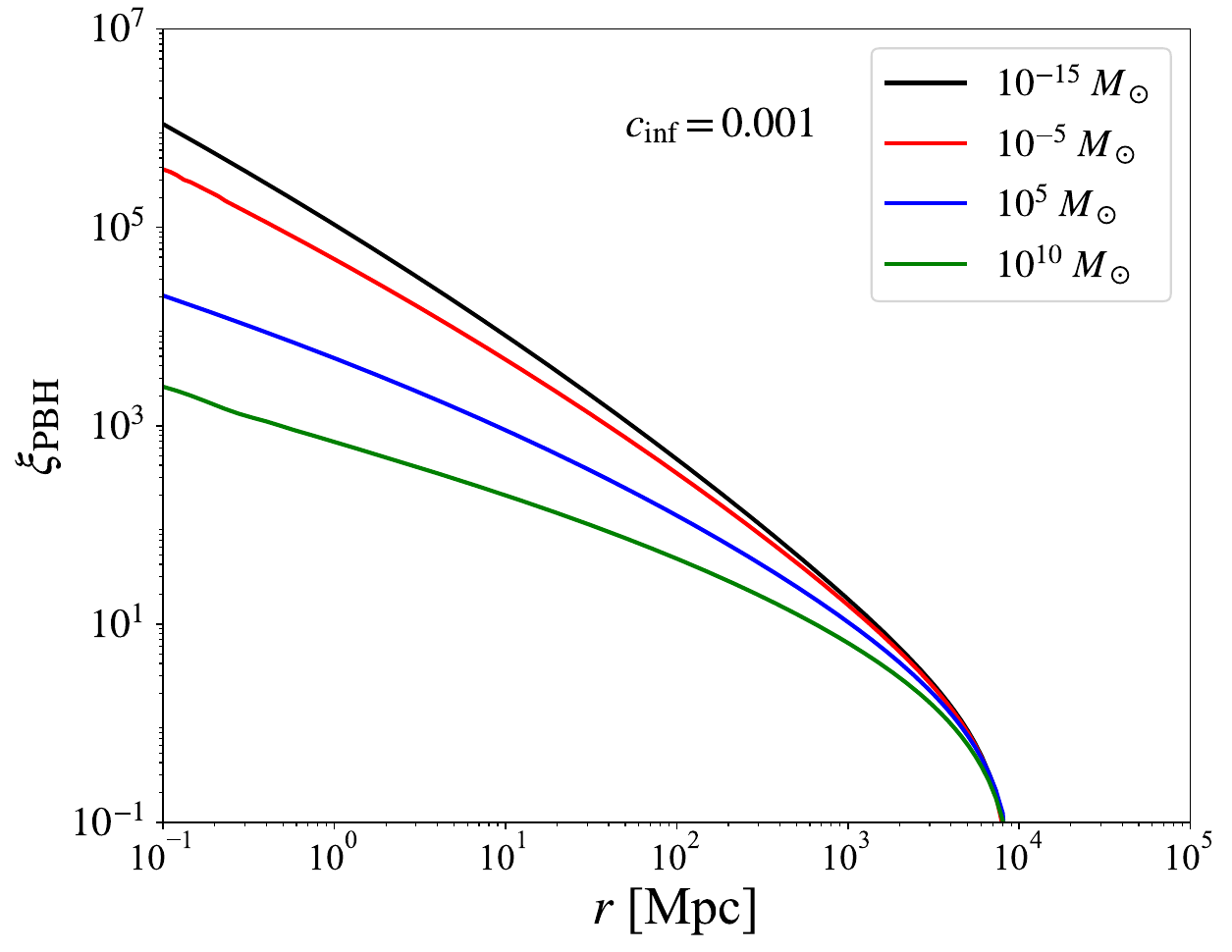} 
\caption{The two-point correlation function of PBHs $\xi_\pb$ as a function of $r$. The left panel shows the dependence on the spectator field mass parameter $c_{\rm inf}$ for fixed $M_{\rm PBH}=1\,M_\odot$, while the right panel shows the dependence on the PBH mass $M_{\rm PBH}$ for fixed $c_{\rm inf}=0.001$. In the left panel, $\xi_{\rm PBH}$ is nearly insensitive to $c_{\rm inf}$ for $c_{\rm inf}\lesssim0.01$, whereas the larger value $c_{\rm inf}=0.1$ produces a substantially weaker clustering. In the right panel, increasing $M_{\rm PBH}$ systematically suppresses $\xi_{\rm PBH}$. } \label{fig:xirhoM}
\end{figure}

Finally, the two-point correlation function of PBHs $\xi_{\rm PBH}(r)$ can be obtained from the ratio of the joint probability to the square of the one-point probability as~\cite{Shinohara:2021psq}
\begin{align}
\xi_{\rm PBH}(r) =\f{P_2(r)}{P_1^2}-1. 
\end{align}
When $\xi_{\rm PBH}>0$, the PBHs can be regarded as being clustered. Fig.~\ref{fig:xirhoM} illustrates $\xi_{\rm PBH}$ as a function of $r$ for different PBH masses $M_\pb$ and $c_{\rm inf}$, with the PBH abundance $f_\pb=1$. First, from Eq. \eqref{Pchik}, for $c_{\rm inf}\lesssim0.01$, the power spectrum of the spectator field $\mathcal P_\chi$ is nearly scale-invariant over the relevant range of scales. Consequently, the resulting correlation coefficient $\epsilon$ depends only weakly on $c_{\rm inf}$, so does the PBH correlation function $\xi_{\rm PBH}$ (left panel). In contrast, for $c_{\rm inf}=0.1$, the stronger scale-dependence of $\mathcal P_\chi$ leads to weaker PBH clustering, so the $\xi_\pb$--$r$ relation differs from the former. Second, the PBH correlation function $\xi_{\rm PBH}$ decreases with the PBH mass $M_\pb$ (right panel). This is because a larger $M_\pb$ corresponds to a smaller $k_{\max}$ in Eq. \eqref{kmax}, which reduces the contribution of small-scale modes to the spectator field correlation, leading to a smaller $\epsilon(r)$ and, consequently, a weaker PBH clustering. 

\section{Primordial isocurvature perturbations} \label{sec:iso}

PBH clustering can provide an additional source of density fluctuations beyond the standard adiabatic perturbations. To quantify the impact on the subsequent growth of cosmic structure, we first translate the PBH two-point correlation function $\xi_\pb(r)$ into the corresponding dimensional power spectrum of PBH density fluctuations $P_\pb(k)$. Assuming statistical homogeneity and isotropy, the two-point correlation function of the PBH density contrast $\dt_\pb$ can be written as
\begin{align}
\langle\delta_{\rm PBH}({\bf 0})\delta_{\rm PBH}({\bf x})\rangle=\f{\dt_{\rm D}(\bf x)}{\bar{n}_{\rm PBH}}+\xi_{\rm PBH}(r),
\end{align}
where $\dt_\pb$ is defined as
\begin{align}
\dt_{\rm PBH}({\bf x})=\f{\dt\rho_{\rm PBH}({\bf x})}{\bar\rho_{\rm PBH}}=
\f{1}{\bar{n}_{\rm PBH}}\sum_i\dt_{\rm D}({\bf x}-{\bf x}_i)-1, 
\end{align}
with $\bar{\rho}_{\rm PBH}=\bar{n}_{\rm PBH}M_{\rm PBH}$ being the mean PBH energy density. Therefore, the power spectrum $P_\pb(k)$ is
\begin{align}
P_{\rm PBH}(k)=\int \langle\delta_{\rm PBH}({\bf 0}) \delta_{\rm PBH}({\bf x})\rangle e^{-i{\bf k}\cdot{\bf x}}\,\dd^3 x=P_{\rm Poisson}+P_\xi(k),
\end{align}
where $P_{\rm Poisson} = 1/\bar{n}_{\rm PBH}$ arises from the number fluctuations of the Poisson distributed PBHs, and the clustering contribution is
\begin{align}
P_\xi(k) 
=4\pi\int\tilde{r}^2\f{\sin(k\tilde{r})}{k\tilde{r}}\xi_{\rm PBH}(\tilde{r})\,\dd\tilde{r}, 
\end{align}
which encodes the fluctuations of the PBH number density. These number density fluctuations correspond to an additional matter density perturbation that is initially isocurvature in nature. During the matter-dominated era, this perturbation grows gravitationally and can thus enhance the formation of small-scale structures.

Then, the PBH-induced isocurvature matter power spectrum $P_{\rm iso}$ at redshift $z$ can be obtained from the PBH density fluctuation power spectrum $P_{\rm PBH}(k)$ as~\cite{Inman:2019wvr}
\begin{align}
P_{\rm iso}(k,z) =\left\{
\begin{array}{lr}
[f_{\rm PBH}D_{\rm PBH}(z)]^2 P_{\rm PBH}(k) & \quad (k\leq k_{\rm cut}),\\
0 & \quad (k> k_{\rm cut}),
\end{array}\right.
\end{align}
where
\begin{align}
D_{\rm PBH}(z) \simeq \left(1+\f{3\gamma}{2\alpha_-}\f{1+z_{\rm eq}}{1+z}\right)^{\alpha_-}
\end{align}
is the growth factor of the isocurvature perturbations. Here, $\alpha_-=(\sqrt{1+24\gamma}-1)/4$~\cite{Inman:2019wvr}, and $\gamma=
\Omega_{\rm DM}/\Omega_{\rm m}$, with $\Omega_{\rm m}=0.315$ being the density parameter for total matter~\cite{Planck:2018vyg}. The isocurvature perturbations are restricted to sufficiently large scales, where linear perturbation theory is expected to remain valid. For the initially Poisson distributed PBHs, the cut-off scale is approximately given by the inverse of the mean separation between PBHs, 
\begin{align}
k_{\rm cut}^{\rm Poisson}=(2\pi^2\bar{n}_{\rm PBH})^{1/3}. \label{kcutPoisson}
\end{align}
On the scales corresponding to $k_{\rm cut}^{\rm Poisson}$, the associated comoving volume contains, on average, fewer than one PBH. 
However, if the PBHs are strongly clustered with $\xi_\pb\gg1$, the relevant cut-off scale cannot be handled in the same way as the Poisson case. In this strongly clustered regime, we assume that the PBH-induced isocurvature perturbations dominate over the primary curvature perturbations on the relevant small scales. We define a redshift dependent cut-off scale $k_{\rm cut}^{\rm cl}(z)$ by the condition that the dimensionless isocurvature power spectrum ${\cal P_{\rm iso}}$ indicates the onset of nonlinear structure formation,
\begin{align}
{\cal P_{\rm iso}}(k_{\rm cut}^{\rm cl}(z),z)=\f{[{k_{\rm cut}^{\rm cl}}(z)]^3}{2\pi^2}{P_{\rm iso}(k_{\rm cut}^{\rm cl}(z),z)}=1.
\end{align}
When ${\cal P_{\rm iso}}>1$, the PBH distribution becomes highly nonlinear and the linear power spectrum description breaks down. As a result, we conservatively remove these modes from the variance calculation, which likely underestimates the enhancement of the PBH abundance, since the nonlinear PBH clusters could themselves provide additional gravitational seeds for the subsequent structure formation.

The PBH-induced isocurvature perturbations constitute an additional contribution to the primordial density fluctuations, and thus must be compared with the standard adiabatic curvature perturbations generated during inflation. The power spectrum of primordial curvature perturbation is described by a nearly scale-invariant form as
\begin{align}
P_{\cal R}(k) = \f{2\pi^2}{k^3}A_{\rm s}\lt(\f{k}{k_*}\rt)^{n_{\rm s}-1}, 
\end{align}
with the amplitude $A_{\rm s}=2.1\times10^{-9}$ and the spectral index $n_{\rm s}=0.965$~\cite{Planck:2018vyg}. The CMB observations by the Planck satellite have placed stringent constraints on primordial isocurvature perturbations, since it represents an additional isocurvature mode. The isocurvature fraction at the pivot scale $k_*$ is defined as
\begin{align}
\beta_{\rm iso}(k_*)=\f{P_{\rm iso}(k_*)}{P_{\rm iso}(k_*)+P_{\cal R}(k_*)}.
\end{align}
In this work, we adopt the Planck constraints at the $95 \%$ CL~\cite{Planck:2018jri},
\begin{align}
\beta_{\rm iso}&<0.039\quad (k_*=0.1\,{\rm Mpc}^{-1}),\label{beta1}\\
\beta_{\rm iso}&<0.038\quad (k_*=0.05\,{\rm Mpc}^{-1}),\label{beta05}\\
\beta_{\rm iso}&<0.035\quad (k_*=0.002\,{\rm Mpc}^{-1}).\label{beta002}
\end{align}
The three bounds in Eqs.~\eqref{beta1}--\eqref{beta002} correspond to different
pivot scales and lead to different constraints on the PBH parameter space. 

\begin{figure}[htb]
\centering
\includegraphics[width=0.48\linewidth]{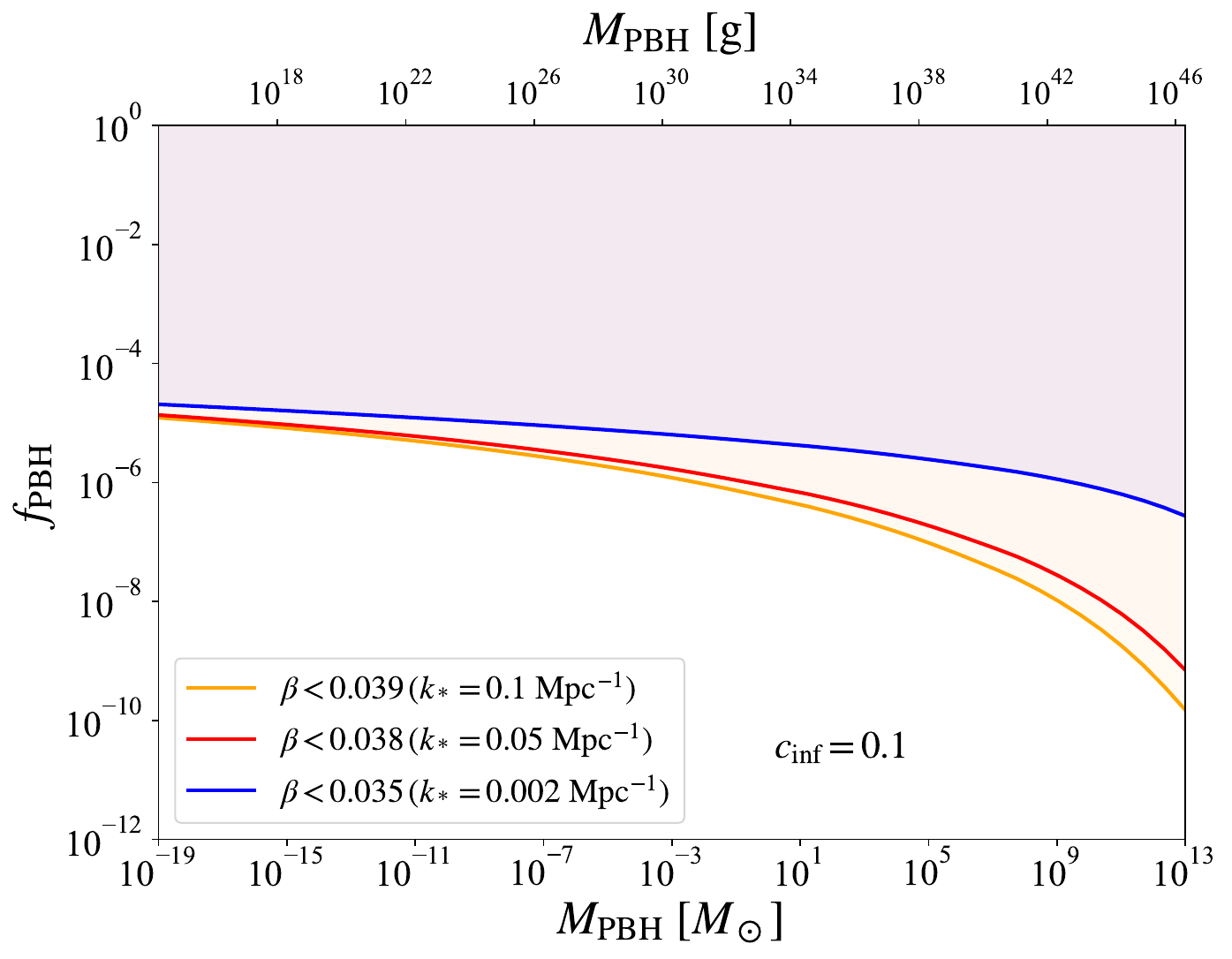} \quad
\includegraphics[width=0.48\linewidth]{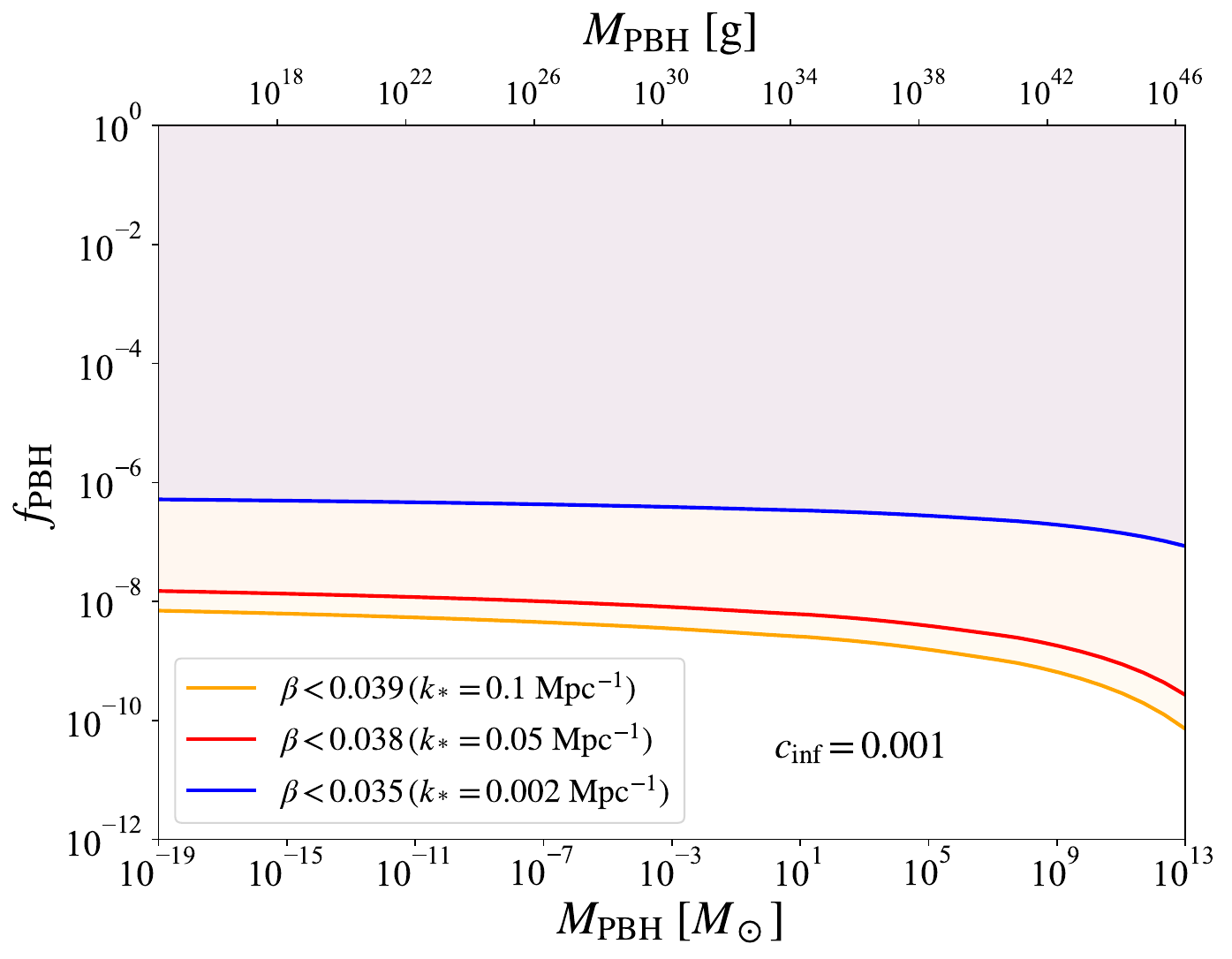}
\caption{The constraints on PBH abundance $f_\pb$ as a function of the PBH mass $M_\pb$ from the CMB observations~\cite{Planck:2018vyg}, with $c_{\rm inf}=0.1$ (left panel) and $c_{\rm inf}=0.001$ (right panel). The three curves correspond to the Planck 95\% CL constraints $\beta_{\rm iso}<0.039$, $0.038$, and $0.035$, evaluated at $k_*=0.1$, $0.05$, and $0.002\,{\rm Mpc}^{-1}$, respectively. The shaded regions above the corresponding constraint curves are excluded. } 
\label{fig:fPBH_limits}
\end{figure}

The constraints on the PBHs arising from the spectator field are shown in Fig.~\ref{fig:fPBH_limits}, with the parameter $c_{\rm inf}=0.1$ and $0.001$, respectively. The upper bound on the PBH abundance $f_{\rm PBH}$ becomes more stringent as the PBH mass $M_{\rm PBH}$ increases. It is evident that the constraint evaluated at $k_*=0.1\,{\rm Mpc}^{-1}$ gives the most restrictive bound on the PBH abundance, whereas the constraint at $k_*=0.002\,{\rm Mpc}^{-1}$ allows a substantially larger region of parameter space. Such a pronounced sensitivity to the choice of the pivot scale naturally originates from the strong scale-dependence of the PBH-induced isocurvature power spectrum $\mathcal{P}_{\rm iso}$. Since the primary CMB temperature anisotropies and polarization data impose the strongest and most robust constraints on large spatial scales (i.e., the small-$k$ regime), we use $k_*= 0.002\,{\rm Mpc}^{-1}$ as the main analysis benchmark.
Furthermore, to assess how sensitive our predictions are to this choice, we compare the results across different pivot scales using Eqs.~\eqref{beta1}--\eqref{beta002} below. This comprehensive comparison allows us to evaluate the robustness of explaining the JWST-inferred stellar mass density via PBH clustering. 

\section{Structural formation enhanced by the PBH clustering}
\label{sec:clustering}

PBH clustering can influence cosmic structure formation through two physically distinct channels, depending on the scale in consideration. On sufficiently large scales with $k_{\min}\leq k\leq k_{\rm cut}^{\rm cl}$, the spatial fluctuations of the PBH distribution source isocurvature perturbations, which subsequently grow through gravitational instability~\cite{Meszaros:1975ef, 1983ApJ...268....1C}. On smaller scales with $k_{\rm cut}^{\rm cl}< k\leq k_{\rm max}$, the PBHs can become gravitationally bound into compact clusters. These nonlinear PBH clusters can then act as localized gravitational seeds, inducing early matter accretion through the seed effect~\cite{1983ApJ...268....1C, Carr:1984ikr, Carr:2018rid}. It is important to distinguish the compact PBH clusters from the PBH clustering. PBH clustering refers to the general non-Poisson spatial distribution of PBHs characterized by a positive two-point correlation function $\xi_{\rm PBH}(r)>0$. Within this clustered distribution, the regions with sufficiently strong clustering can undergo nonlinear gravitational evolution, causing PBHs to become gravitationally bound into compact groups. We refer to these strongly clustered and gravitationally bound groups as compact PBH clusters, which can act as localized gravitational seeds for structure formation. 

In this section, we discuss the enhancement of structure formation from the isocurvature perturbations and the seed effect in detail, and evaluate whether they can explain the JWST observations. 

\subsection{Halo mass function from the isocurvature perturbations}
\label{sec:halo}


When $k_{\min}\leq k\leq k_{\rm cut}^{\rm cl}$, the role of PBHs in our scenario is to modify the small-scale matter distribution through the clustered isocurvature perturbations. We will investigate whether PBH clustering generated by the spectator field can provide an additional channel that enhances early structure formation. 

We first compute the total matter power spectrum including the PBH-induced isocurvature contribution and the standard adiabatic matter fluctuations. The variance of the matter density field smoothed on a mass scale $M_{\rm h}$ is characterized by the total power spectrum $P_{\rm tot}$,
\begin{align}
\sg^2(M_{\rm h},z)&=\int P_{\rm tot}(k,z)W^2(kR)\f{k^2}{2\pi^2}\,\dd k \n\\
&=\int\lt[D^2(z)P_{\rm \Lambda CDM}(k,0) +P_{\rm iso}(k,z)\rt] W^2(kR)\f{k^2}{2\pi^2} \,\dd k, 
\end{align}
where $W(kR)=3[\sin (kR)-kR\cos(kR)]/(kR)^3$ is the window function in coordiante space, $R=[3M_{\rm h}/(4\pi\rho_{\rm m})]^{1/3}$ is the relevant comoving radius, $P_{\rm \Lambda CDM}(k)$ denotes the adiabatic matter power spectrum for the $\Lambda$CDM model without PBHs, and $D(z)\propto H(z)\int_z^\infty(1+\tilde{z})/H(\tilde{z})^3\,\dd \tilde{z}$ is the linear growth factor, with the normalization as $D(0)=1$. Below, we evaluate $P_{\rm \Lambda CDM}(k)$ by using \texttt{CAMB}~\cite{Lewis:1999bs} and approximate the growth factor at $z\geq 1$ as~\cite{Iocco:2024rez} 
\begin{align}
D(z)\simeq\f{1.27}{1+z}. 
\end{align}

We determine the halo mass function ${\dd n_{\rm h}}/{\dd M_{\rm h}}$ by using the Sheth--Tormen prescription~\cite{Sheth:1999mn, Sheth:1999su}, which extends the original Press--Schechter formalism~\cite{Press:1973iz}. In this framwork, the halo mass function can be written as 
\begin{align}
\f{\dd n_{\rm h}}{\dd M_{\rm h}} =\f{\rho_{\rm m}}{M_{\rm h}}f(\nu_{\rm c}(M_{\rm h},z)) \lt|\f{\dd\nu_{\rm c}(M_{\rm h},z)}{\dd M_{\rm h}}\rt|, 
\end{align}
where the peak height is characterized as
\begin{align}
\nu_{\rm c}(M_{\rm h},z)=\f{\dt_{\rm c}}{\sg(M_{\rm h},z)}, 
\end{align}
and
\begin{align}
f(\nu_{\rm c})=\sqrt{\f{2q}{\pi}}\f{1}{1+\Gamma(1/2-p)/ (2^p\sqrt{\pi})}\lt[1+\f{1}{(q\nu_{\rm c}^2)^p}\rt] \exp\lt(-\f{q\nu_{\rm c}^2}{2}\rt). 
\end{align}
Throughout our analysis, we take the parameters as $p=0.3$ and $q=0.85$~\cite{Schneider:2020xmf}. 

Below, we assess the viability of initially clustered PBHs as a possible origin of the massive galaxy population observed by the JWST. Our approach is to confront the predicted number density of massive galaxies in our PBH model with the latest JWST measurements. As a benchmark, we adopt the inferred stellar masses $M^{\rm obs}_*$ spanning $10^9$--$10^{11}\,M_\odot$ over the redshift interval $7\leq z\leq10$ reported by the CEERS program~\cite{Labbe:2022ahb}. At $z=8$, the cumulative stellar mass density associated with the galaxies satisfying $M_*^{\rm obs}\geq10^{10}\,M_\odot$ is~\cite{Labbe:2022ahb, Liu:2022bvr}
\begin{align}
\rho_{\rm obs}(M_* \geq 10^{10}\,M_\odot)\simeq1.3\times10^6\,M_\odot\,{\rm Mpc}^{-3}.
\end{align}
To connect the predicted halo abundance with observations, we next map the halo mass onto the stellar mass by using a simple star formation prescription. Here, we speculate that every DM halo contains a single central galaxy, so the relation between the halo and total stellar masses is 
\begin{align}
M_{\rm h}(M_*) = \f{M_*}{f_{\rm b}\ep_*}, \label{Mh}
\end{align}
where $f_{\rm b}
=0.157$ is the baryon fraction~\cite{Planck:2018vyg}, and $\ep_*$ is the star formation efficiency (not to be confused with the correlation coefficient $\ep$). Then, the expected number density of galaxies with $M_*\geq10^{10}\,M_\odot$ is given by
\begin{align}
n_{\rm gal}(M_*\geq 10^{10}\,M_\odot) =\int_{M_{\rm h}(M_{*}=10^{10}\,M_\odot)}^\infty\f{\dd n_{\rm h}(M_{\rm h},z)}{\dd M_{\rm h}}\,\dd M_{\rm h},
\end{align}
where we take $z=8$, and the comoving cumulative stellar mass density is
\begin{align}
\rho(M_*\geq 10^{10}\,M_\odot)=f_{\rm b}\ep_*\int_{M_{\rm h}(M_{*}=10^{10}\,M_\odot)}^\infty M_{\rm h}\f{\dd n_{\rm h}(M_{\rm h},z)}{\dd M_{\rm h}}\,\dd M_{\rm h}.
\end{align}

\begin{figure}[htb]
\centering
\includegraphics[width=0.48\textwidth]{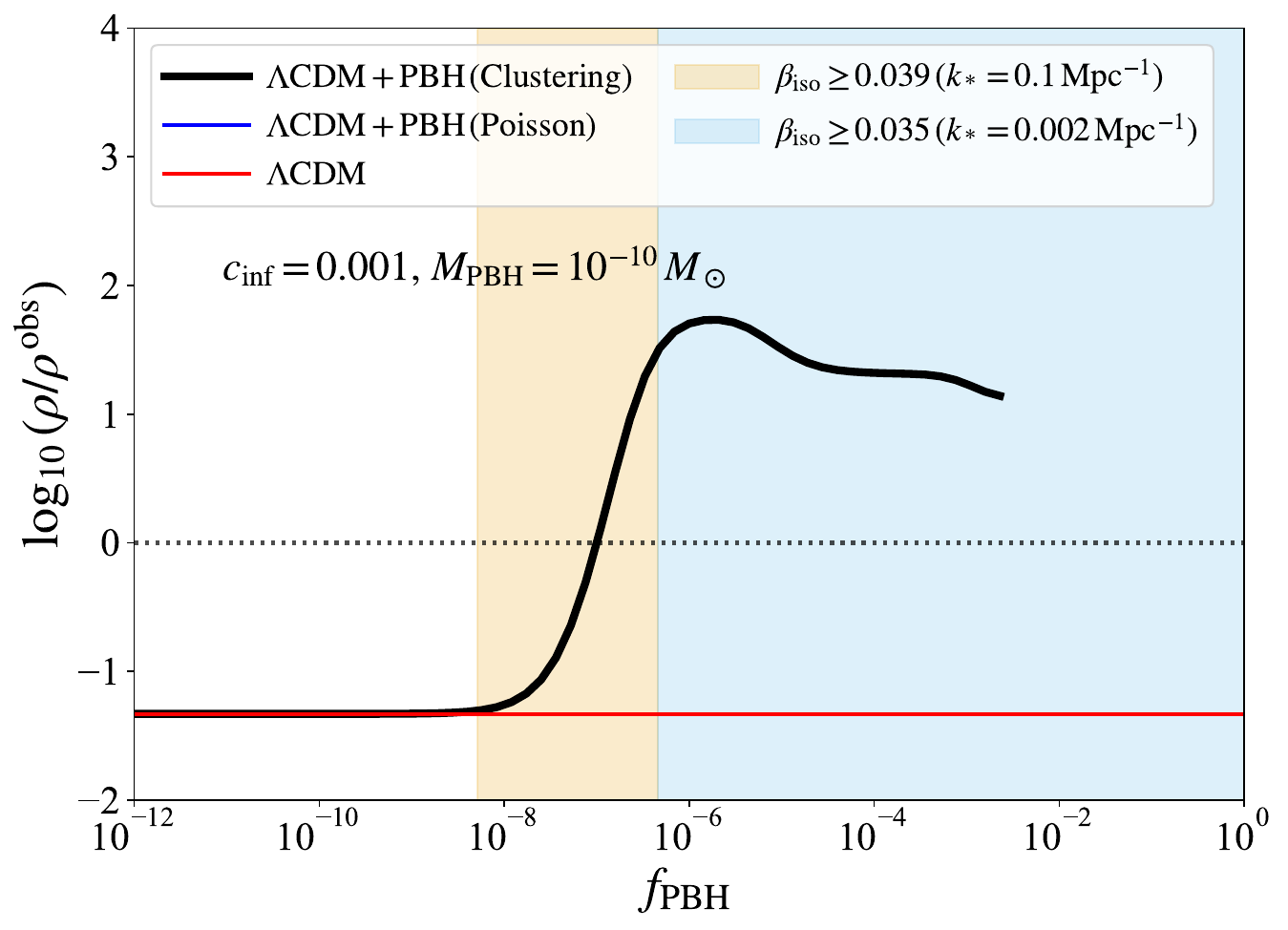}\quad
\includegraphics[width=0.48\textwidth]{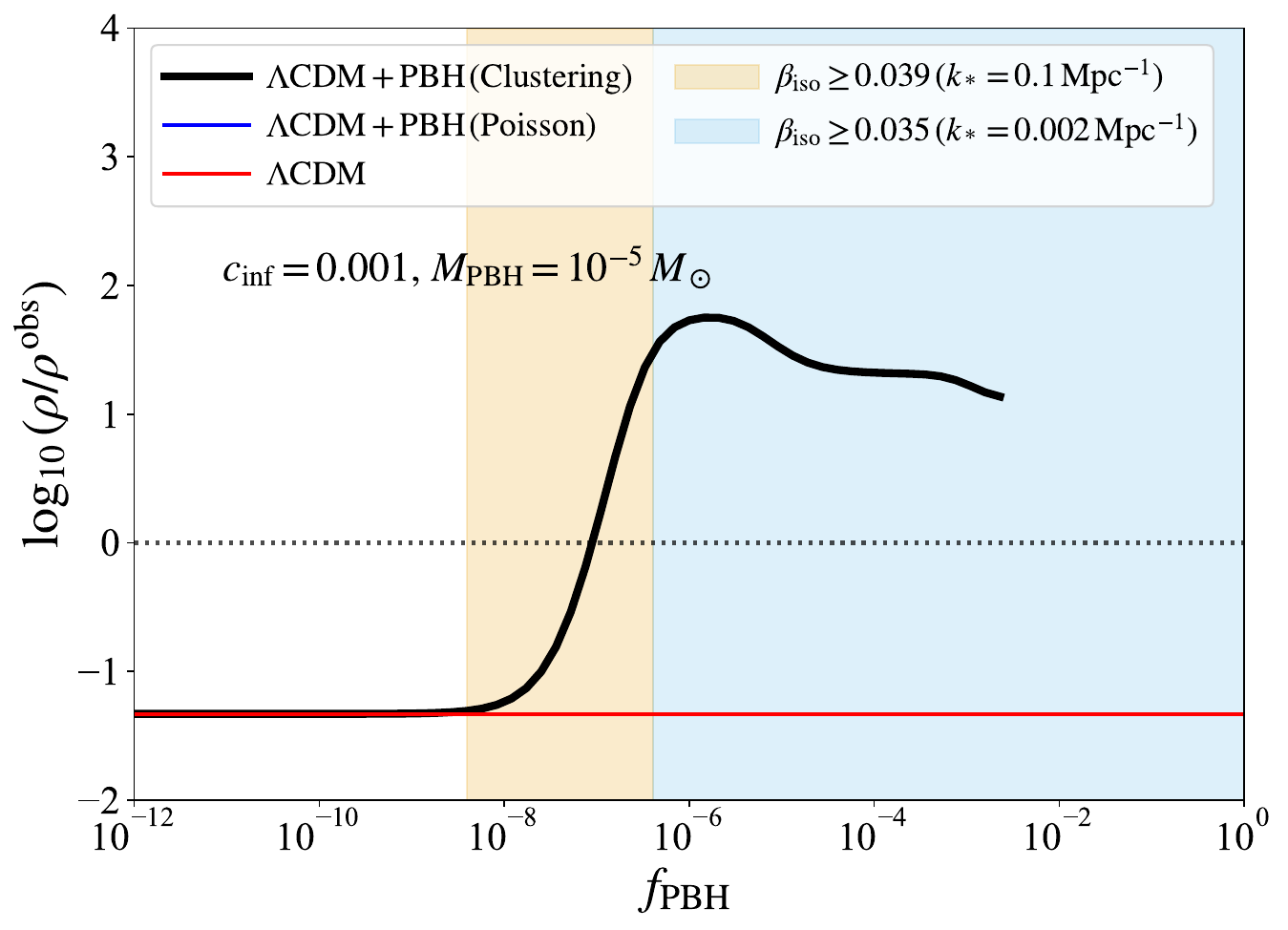} \\
\includegraphics[width=0.48\textwidth]{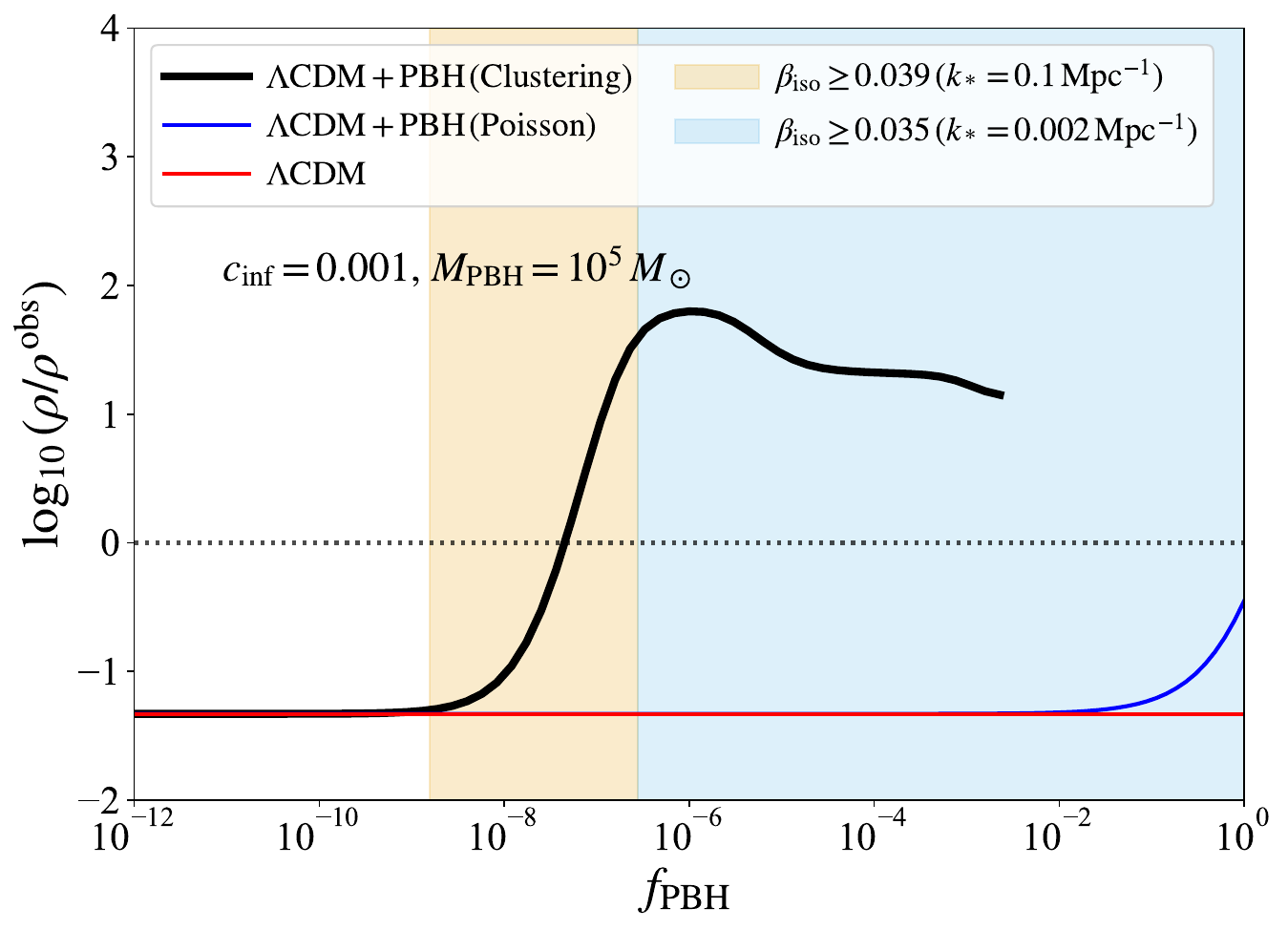}\quad
\includegraphics[width=0.48\textwidth]{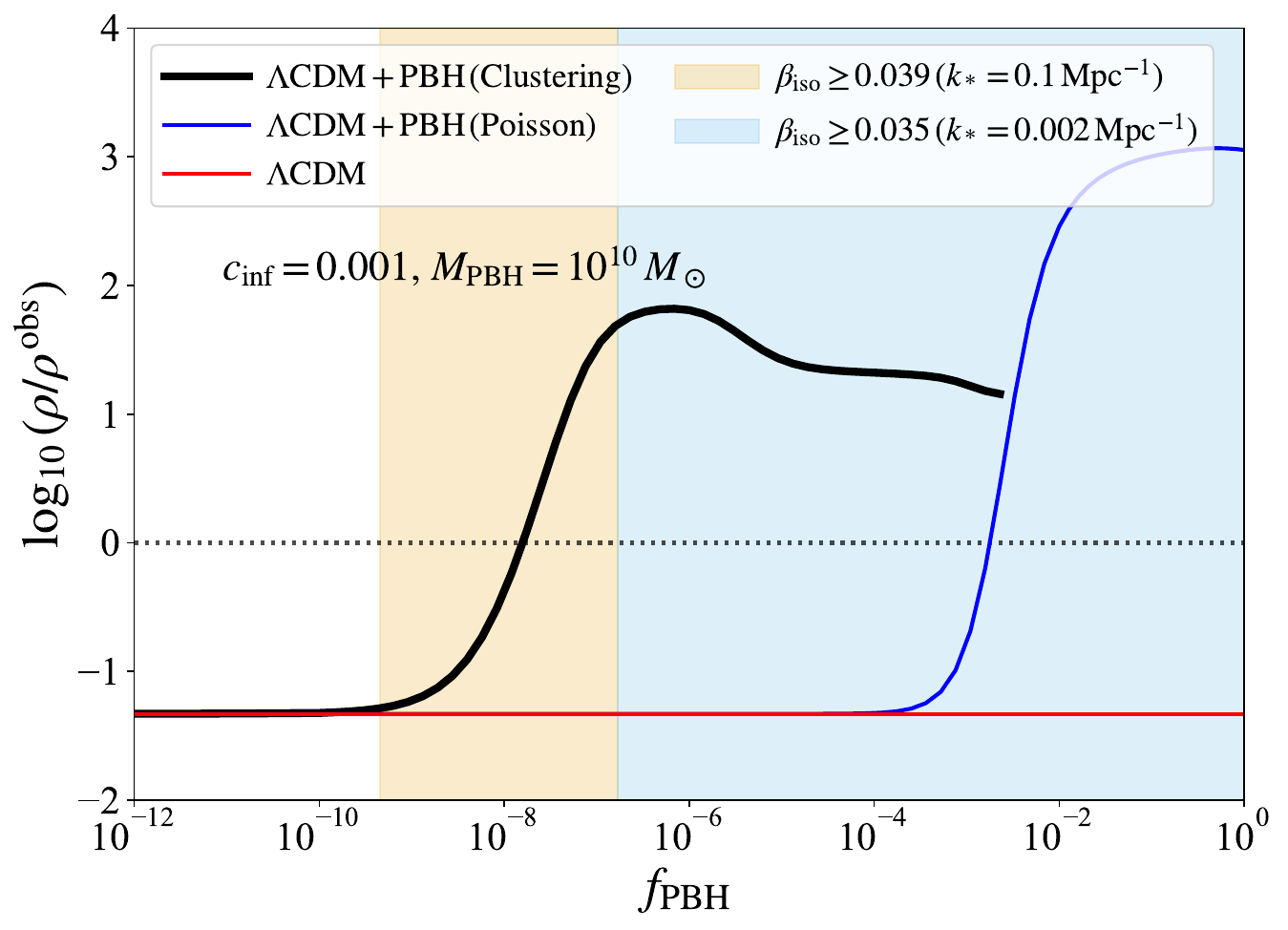}
\caption{The comoving cumulative stellar mass density $\rho/\rho^{\rm obs}$ arising from the isocurvature perturbations as a function of the PBH abundance $f_\pb$, with the parameter $c_{\rm inf}=0.001$ and different PBH masses $M_\pb=10^{-10}$, $10^{-5}$, $10^5$, and $10^{10}~M_\odot$. The black, blue, and red lines represent the cases of $\Lambda{\rm CDM} + {\rm PBH~(Clustering)}$, $\Lambda{\rm CDM} + {\rm PBH~(Poisson)}$, and $\Lambda$CDM only, respectively. The horizontal dotted line corresponds to $\rho/\rho^{\rm obs}=1$, indicating that the predicted comoving cumulative stellar mass density matches the value inferred from the JWST observations. The blue shaded region is excluded by the CMB isocurvature constraint on the benchmark pivot scale $k_*=0.002\,{\rm Mpc}^{-1}$ in Fig.~\ref{fig:fPBH_limits}, while the orange shaded region indicates the corresponding excluded region for $k_*=0.1\,{\rm Mpc}^{-1}$, illustrating the dependence of the constraint on the choice of the pivot scale. The star formation efficiency is fixed as $\epsilon_*=0.3$. In the $\Lambda$CDM case, the resulting stellar mass density is insufficient to reproduce the value inferred from the JWST. For the Poisson distributed PBHs, reproducing the JWST-inferred stellar mass density requires extremely massive PBHs with $M_{\rm PBH}\gtrsim 1.79\times10^5\,M_\odot$ and the large PBH abundance that have been jointly ruled out by existing observations~\cite{Carr:2020gox, Murgia:2019duy, 1999ApJ...516..195C, Inoue:2017csr}. In contrast, PBH clustering can significantly enhance early structure formation, allowing the JWST-inferred stellar mass density to be reproduced, while remaining consistent with the benchmark isocurvature constraint at $k_*=0.002\,{\rm Mpc}^{-1}$. } 
\label{fig:ab_vs_fPBH_00}
\end{figure}

Finally, adopting the star formation efficiency $\epsilon_*=0.3$ and the parameter $c_{\rm inf}=0.001$, we calculate the comoving cumulative stellar mass density $\rho/\rho^{\rm obs}$ in the PBH clustering scenario. When $\rho/\rho^{\rm obs}=1$, the predicted comoving cumulative stellar mass density can reach the observational value of the JWST. We use the isocurvature fraction constraint $\beta_{\rm iso}<0.035$ at $k_*=0.002\,{\rm Mpc}^{-1}$ as our benchmark, and assess the dependence on the pivot scale by including the constraint $\beta_{\rm iso}<0.039$ at $k_*=0.1\,{\rm Mpc}^{-1}$. Our results are shown in Fig.~\ref{fig:ab_vs_fPBH_00}, together with the predictions for the Poisson distributed PBHs and the $\Lambda$CDM model for comparison. For PBH clustering, as $f_{\rm PBH}$ increases, $\rho/\rho^{\rm obs}$ initially rises significantly, but then slightly drops, because a smaller $k_{\rm cut}^{\rm cl}$ cuts off the small-scale halo formation. When $f_\pb$ is relatively large, the dimensionless power spectrum on the horizon scale satisfies ${\cal P_{\rm iso}}(k_{\rm min})>1$, rendering the linear perturbation framework invalid. Thus, we exclude these regions from our analysis, which can be seen from the truncation on the black solid line. It is worth noting that, the comoving cumulative stellar mass density $\rho/\rho^{\rm obs}$ can reach 1 with different PBH masses even for those well below the solar mass scale, while remaining consistent with the isocurvature fraction constraint at $k_*=0.002\,{\rm Mpc}^{-1}$ in Fig.~\ref{fig:fPBH_limits}. This indicates that the observed stellar mass density inferred from the JWST can be fully reproduced. However, when we consider the isocurvature fraction at $k_*=0.1\,{\rm Mpc}^{-1}$, the allowed enhancement from PBH clustering is strongly suppressed, showing strong dependence on the choice of pivot scale. In contrast, for Poisson distributed PBHs, it is much more difficult to enhance the structure formation. In this case, the JWST observations can be explained only if the PBH mass is extremely large as $M_{\rm PBH}\gtrsim 1.79\times10^5\,M_\odot$, with high PBH abundance $f_{\rm PBH}$, which has been jointly ruled out by observations~\cite{Carr:2020gox, Murgia:2019duy, 1999ApJ...516..195C, Inoue:2017csr}. Furthermore, for the $\Lambda$CDM model, we obtain $\rho/\rho^{\rm obs}=4.78\times10^{-2}$ (i.e., $\log_{10}\rho/\rho^{\rm obs}=-1.33$) for $\epsilon_*=0.3$, which is insufficient to account for the stellar mass density inferred from the JWST observations. 

\begin{figure}[htb]
\centering
\includegraphics[width=0.48\textwidth]{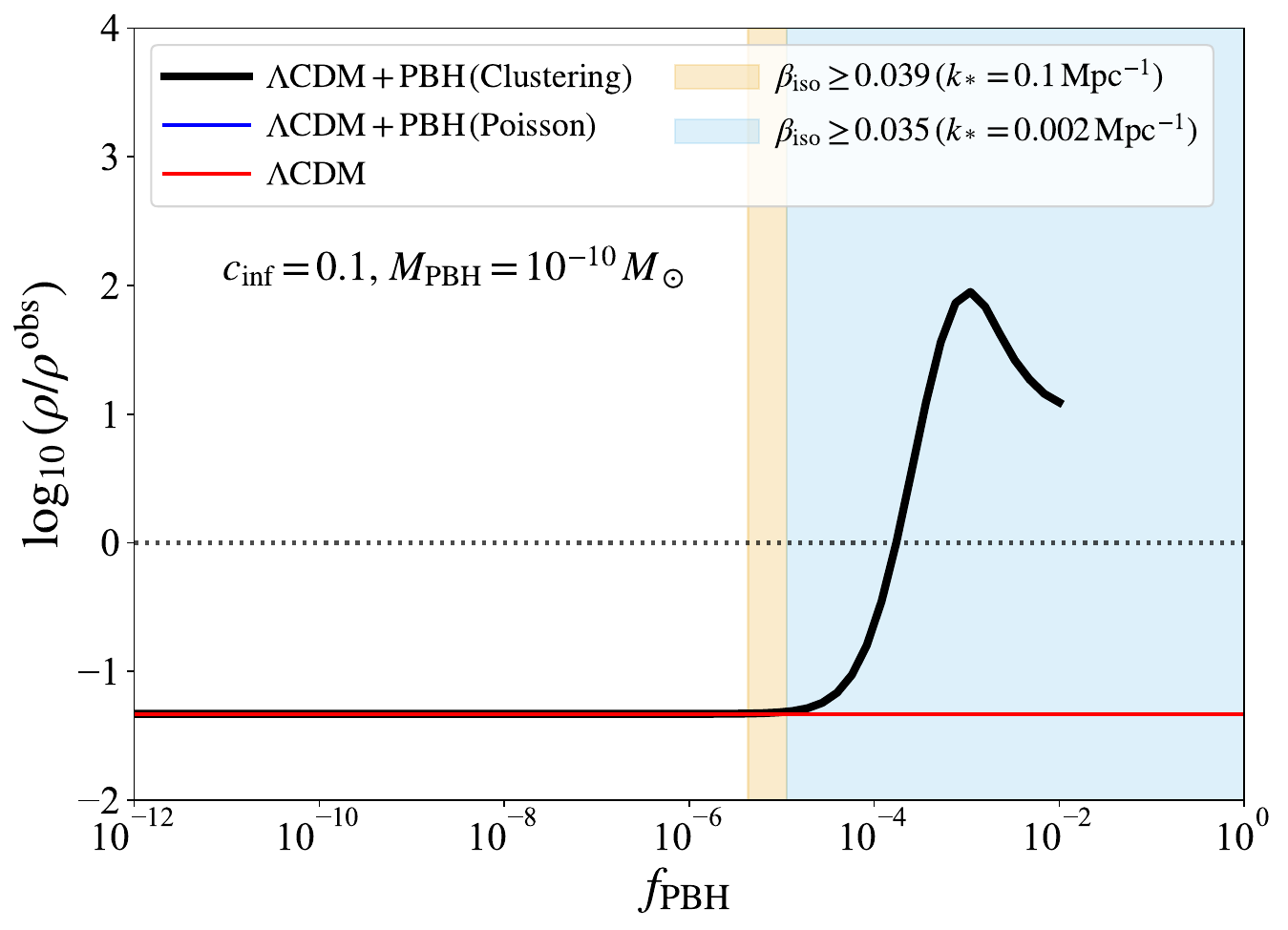}\quad
\includegraphics[width=0.48\textwidth]{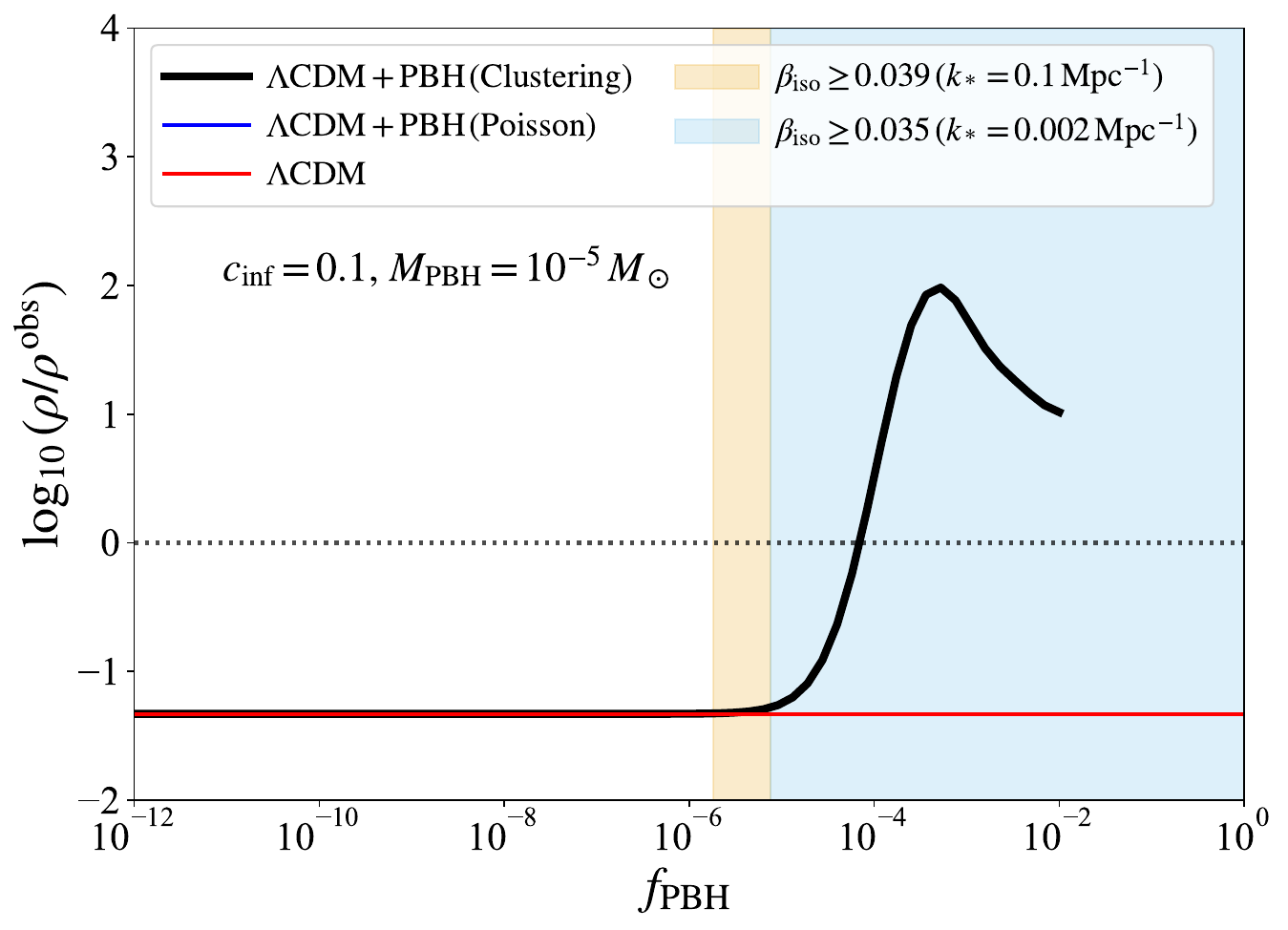}\\
\includegraphics[width=0.48\textwidth]{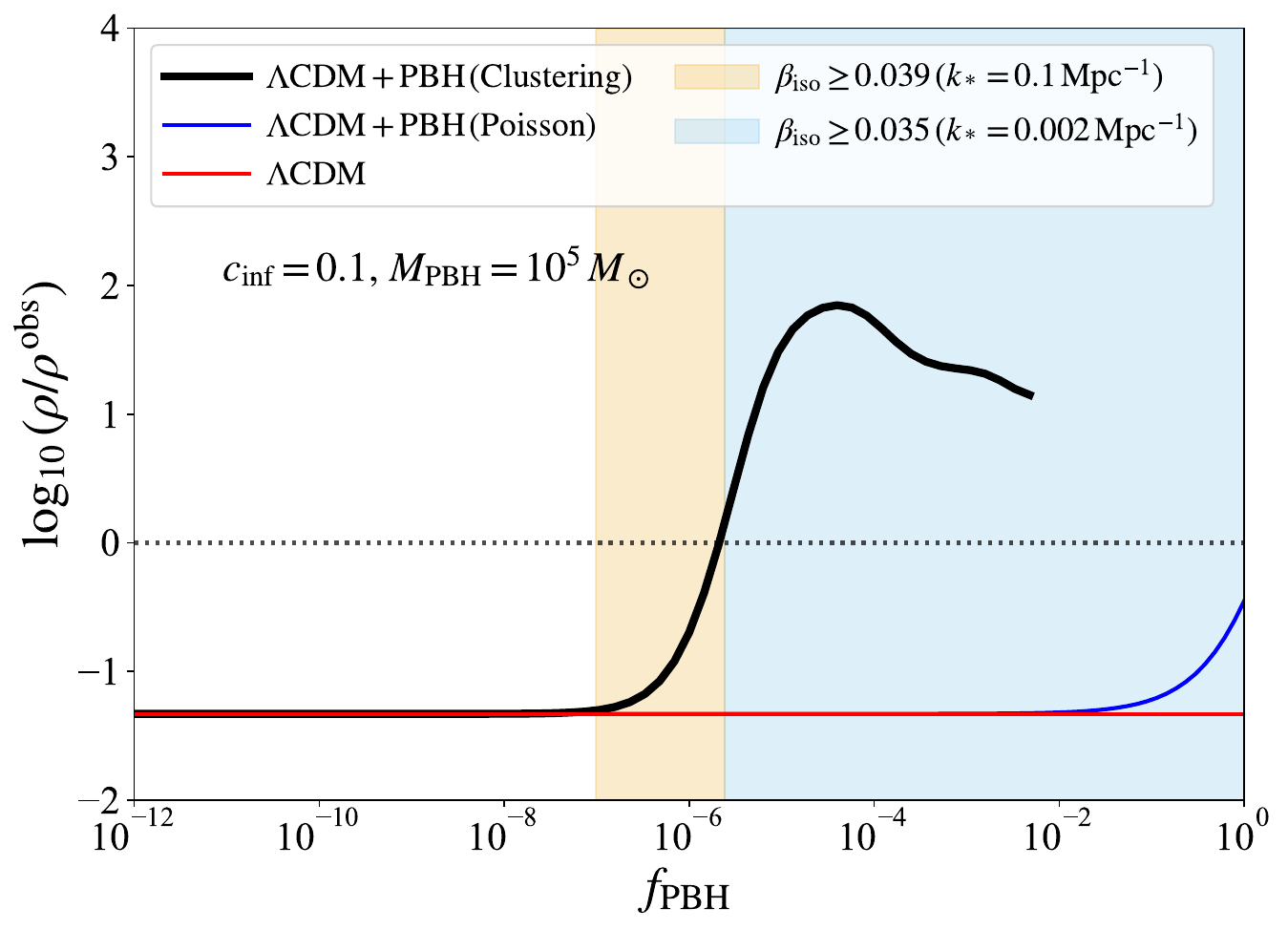}\quad
\includegraphics[width=0.48\textwidth]{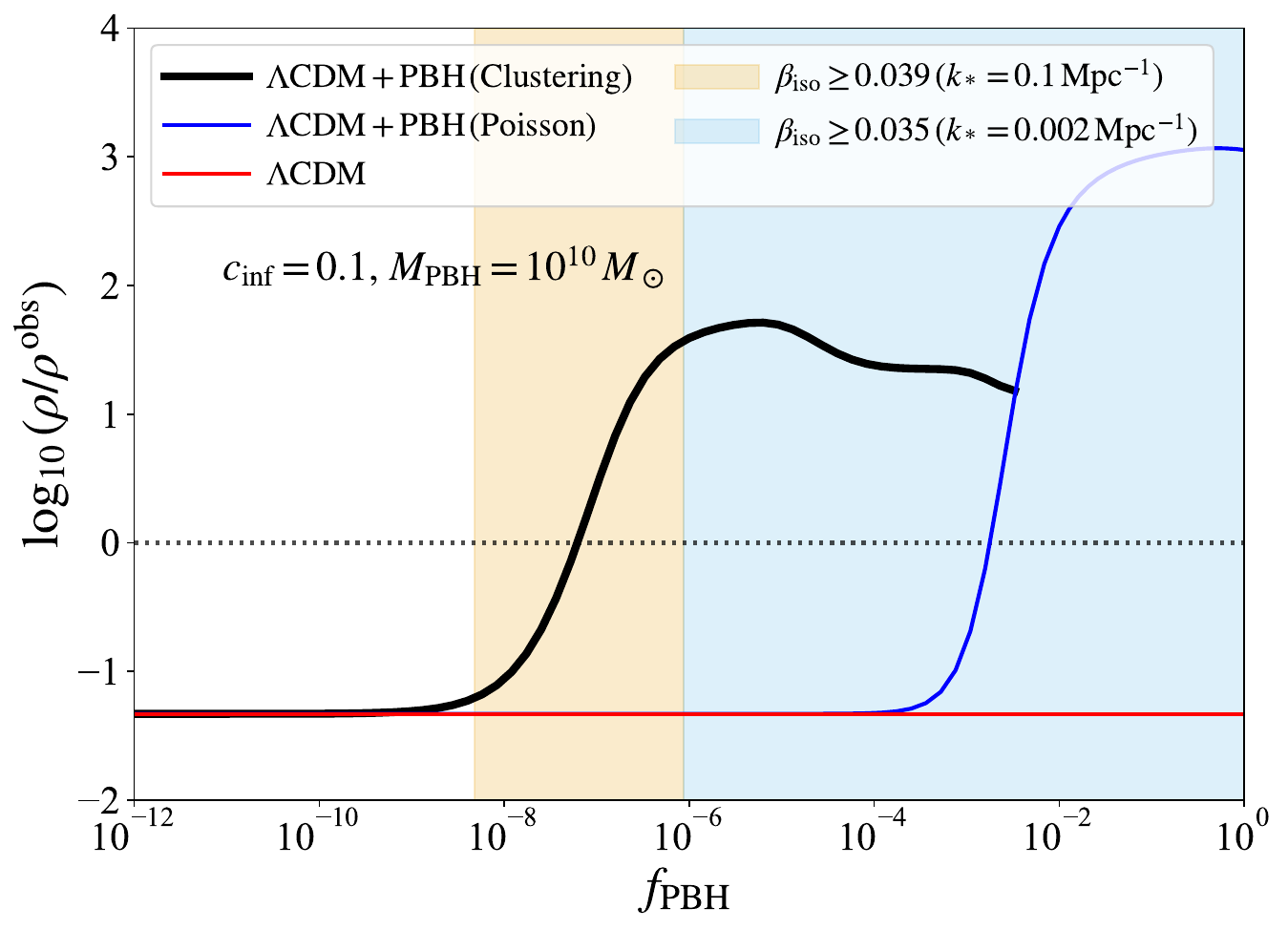}
\caption{Same as Fig.~\ref{fig:ab_vs_fPBH_00}, but with the parameter $c_{\rm inf}=0.1$. For PBH clustering, the resulting stellar mass density depends strongly on the PBH mass $M_\pb$. Significant enhancement of early structure formation while remaining consistent with the benchmark isocurvature constraint at $k_*=0.002\,{\rm Mpc}^{-1}$ is achieved only for $M_\pb\gtrsim10^5\,M_\odot$. } 
\label{fig:ab_vs_fPBH}
\end{figure}

Moreover, when we take parameter $c_{\rm inf}=0.1$, the contribution of PBH clustering varies significantly with the PBH mass $M_\pb$, as shown in Fig.~\ref{fig:ab_vs_fPBH}. Accounting for $\rho/\rho^{\rm obs}=1$, the PBH clustering scenario remains consistent with the CMB isocurvature constraint only for $M_\pb\gtrsim10^5\,M_\odot$. 

\subsection{Galaxy seeds from the compact PBH clusters} \label{sec:seeds}

When $k_{\rm cut}^{\rm cl}< k \leq k_{\max}$, we focus exclusively on the seed mechanism from PBH clusters and neglect the enhancement of the matter power spectrum discussed previously. This allows us to assess separately whether the seed effect of clustered PBHs can account for the massive galaxies observed by the JWST. 

The seed effect arises from the gravitational growth of sufficiently compact PBH clusters after matter--radiation equality. Although PBHs are spatially correlated already at formation in the spectator field scenario, during the radiation-dominated era, their gravitational growth is suppressed by the dominant radiation background. After matter--radiation equality, PBHs can cluster gravitationally, allowing sufficiently overdense PBH clusters to act as localized seeds for the subsequent accretion of surrounding matter~\cite{Carr:2018rid}. The initial density perturbations associated with the seed of mass $M_{\rm seed}$ embedded in a region of total mass $M_{\rm h}$ is approximately
\begin{align}
\delta_{\rm ini}\simeq \frac{M_{\rm seed}}{M_{\rm h}}.
\end{align}
After matter--radiation equality, the seed-induced perturbation grows through gravitational instability, leading to a characteristic halo mass at redshift $z$~\cite{Carr:2018rid},
\begin{align}
M_{\rm h}\simeq \frac{1+z_{\rm eq}}{1+z}M_{\rm seed},\label{Mhseed}
\end{align}
where $z_{\rm eq}\simeq3400$ is the redshift of matter--radiation equality. From Eqs.~\eqref{Mh} and \eqref{Mhseed}, we can determine $M_{\rm seed}$. For isolated PBHs, the seed mass is just the PBH mass $M_{\rm seed}=M_{\rm PBH}$. Requiring such a seed to grow into a halo massive enough to host a galaxy implies $M_{\rm PBH}\gtrsim10^4\,M_\odot$, consistent with previous studies on the PBH seed mechanism~\cite{Carr:2018rid}. 

Since all PBHs are assumed to have the same mass, the fluctuation in PBH number density is equivalent to that in PBH mass density. Due to the isotropy of space, the variance of PBH number fluctuations within a comoving sphere of radius $R$ is defined as
\begin{align}
\sigma_{\rm cl}^2(\bar{N})=\f{\langle(N-\bar{N})^2\rangle}{\bar{N}^2}=\f{1}{\bar{N}} + \f{4\pi}{V_R^2} \int_0^{\min(2R,2r_{\max})} r^2 V_{\rm int}(r,R)\xi_{\rm PBH}(r)\,\dd r,\label{sgclV}
\end{align}
where $r_{\max}=\pi/k_{\rm cut}^{\rm cl}$ defines the physical boundary of an individual nonlinear cluster, $V_{\rm int}(r,R)$ represents the volume of the overlapping part of two spheres with a radius $R$ and a center distance $r$, $V_R= 4\pi R^3/3$ is the volume of the sphere with the comoving radius $R$, and $\bar{N}=4\pi R^3\bar{n}_\pb/3$ is the mean number of PBHs within $V_R$. According to the geometric relationship, the analytical expression for this overlapping volume is
\begin{align}
V_{\rm int}(r,R)=\left\{
\begin{array}{ll}
V_R\lt(1-\f{3r}{4R}+\f{r^3}{16R^3}\rt) & \quad (0\leq r\leq 2R),\\
0 & \quad (r>2R).
\end{array}\right. \label{Vint}
\end{align}
With Eqs.~\eqref{sgclV} and \eqref{Vint}, we obtain
\begin{align}
\sigma_{\rm cl}^2(\bar{N})=\f{1}{\bar{N}} + \f{3}{R^3} \int_0^{2R}r^2 \lt(1-\f{3r}{4R}+\f{r^3}{16R^3}\rt)\xi_{\rm PBH}(r)\,\dd r. \label{sgcl}
\end{align}
When the PBH abundance $f_\pb\to0$, the spatial correlations $\xi_{\rm PBH}(r)\to0$ accordingly. At this point, the variance of PBH number fluctuations $\sigma_{\rm cl}^2(\bar{N})\to{1}/{\bar{N}}$, and the seed effect is governed purely by isolated supermassive PBHs. 

To evaluate the contribution of the seed effect on nonlinear scales, we employ the Press--Schechter formalism~\cite{Press:1973iz}, instead of the Sheth--Tormen model used in Sec.~\ref{sec:halo}. While the Sheth--Tormen prescription incorporates ellipsoidal collapse and tidal shear effects, the Press--Schechter approach offers a direct estimate of local halo formation in compact PBH clusters. Similar to Sec.~\ref{sec:halo}, the peak significance is
\begin{align}
\nu_{\rm c}^{\rm cl}(\bar{N}, z)=\f{\dt_{\rm c}}{\sigma_{\rm cl}(\bar{N},z)}=\f{\dt_{\rm c}}{D(z)\sigma_{\rm cl}(\bar{N})}.
\end{align}
According to the Press--Schechter halo mass function, the mean differential comoving number density of PBH clusters $n_{\rm cl}$ with the mean PBH number $\bar{N}$ is
\begin{align}
\frac{\dd n_{\rm cl}}{\mathrm{d}\bar{N}} = \sqrt{\frac{2}{\pi}} \frac{\bar{n}_{\rm PBH}}{\bar{N}}\left| \frac{\dd\nu_{\rm c}^{\rm cl}(\bar{N}, z)}{\dd\bar{N}} \right| \exp\left\{-\frac{[\nu_{\rm c}^{\rm cl}(\bar{N}, z)]^2}{2} \right\}. \label{SP}
\end{align}
Then, the expected number density of galaxies with $M_*>10^{10}\,M_\odot$ is
\begin{align}
n_{\rm gal}(M_*\geq10^{10}\,M_\odot)=\int_{N_{\rm min}}^{\infty} \frac{\mathrm{d}n_{\rm cl}(\bar{N}, z)}{\mathrm{d}\bar{N}}\,\dd\bar{N},
\end{align}
where $N_{\rm min}$ is determined by the seed mass as $N_{\min}={M_{\rm seed}}/{M_{\rm PBH}}$.
To ensure that the PBH clusters remain spatially isolated and do not overlap with each other, we further demand
\begin{align}
n_{\rm cl} < \frac{3}{4\pi r_{\max}^3}. 
\end{align}
When $f_{\rm PBH}$ is sufficiently high, such that $n_{\rm cl}$ exceeds the exclusion threshold, the isolated seed assumption becomes invalid. Instead of ceasing to produce galaxies, the overlapping clusters would generate additional large-scale fluctuations in the PBH cluster number density, which could in principle contribute to the isocurvature-driven growth of structure. However, we do not include this contribution in our present analysis, because the corresponding parameter region is already strongly constrained by the CMB observations, as shown in Fig.~\ref{fig:fPBH_limits}. Finally, the comoving cumulative stellar mass density from the seed effect reads
\begin{align}
\rho(M_*\geq 10^{10}\,M_\odot)=10^{10}\,M_\odot\int_{N_{\rm min}}^\infty \frac{\bar{N}}{N_{\min}}\frac{\dd n_{\rm cl}(\bar{N},z)}{\dd\bar N}\,\dd \bar{N}.
\end{align}

\begin{figure}[htb]
\centering
\includegraphics[width=0.48\textwidth]{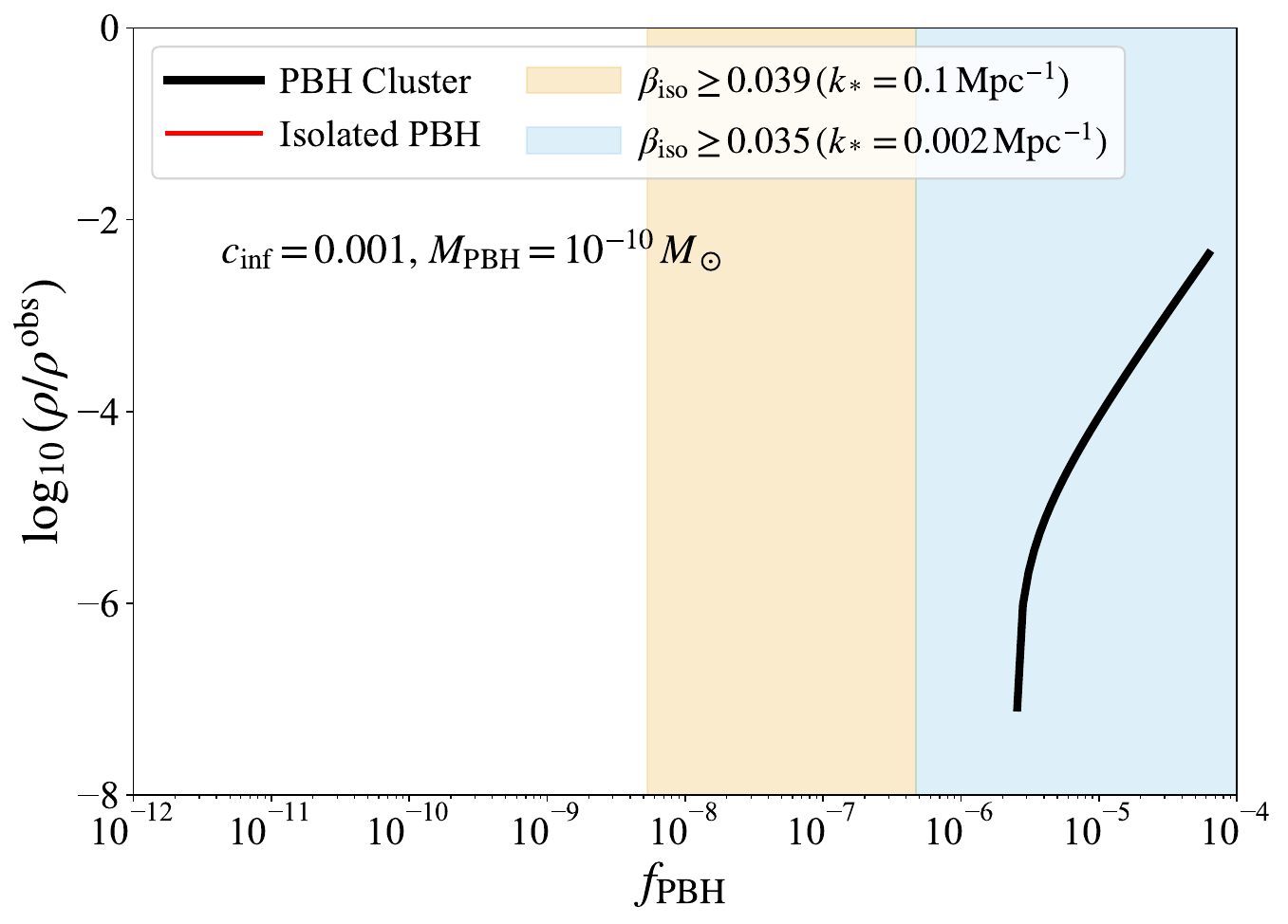}\quad
\includegraphics[width=0.48\textwidth]{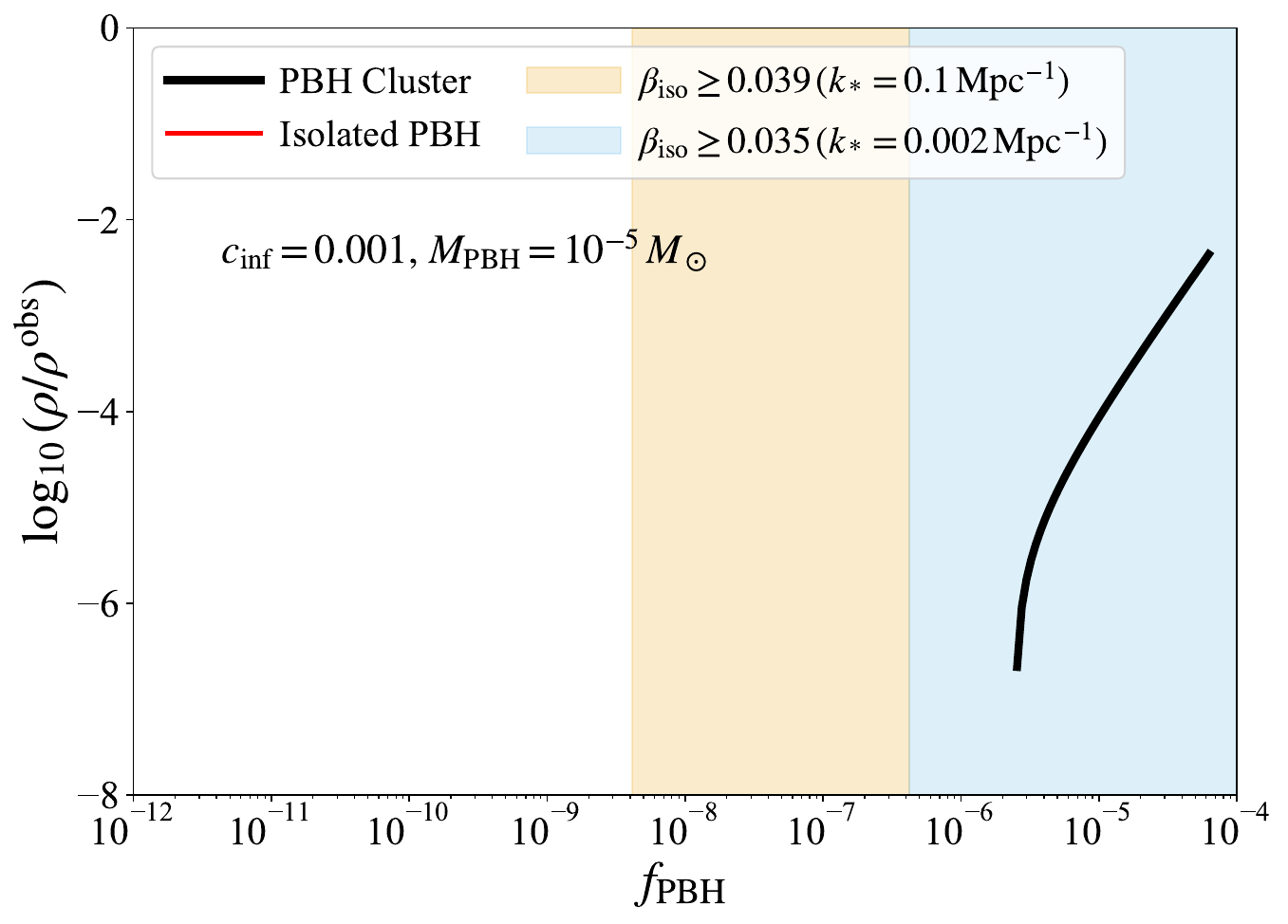}\\
\includegraphics[width=0.48\textwidth]{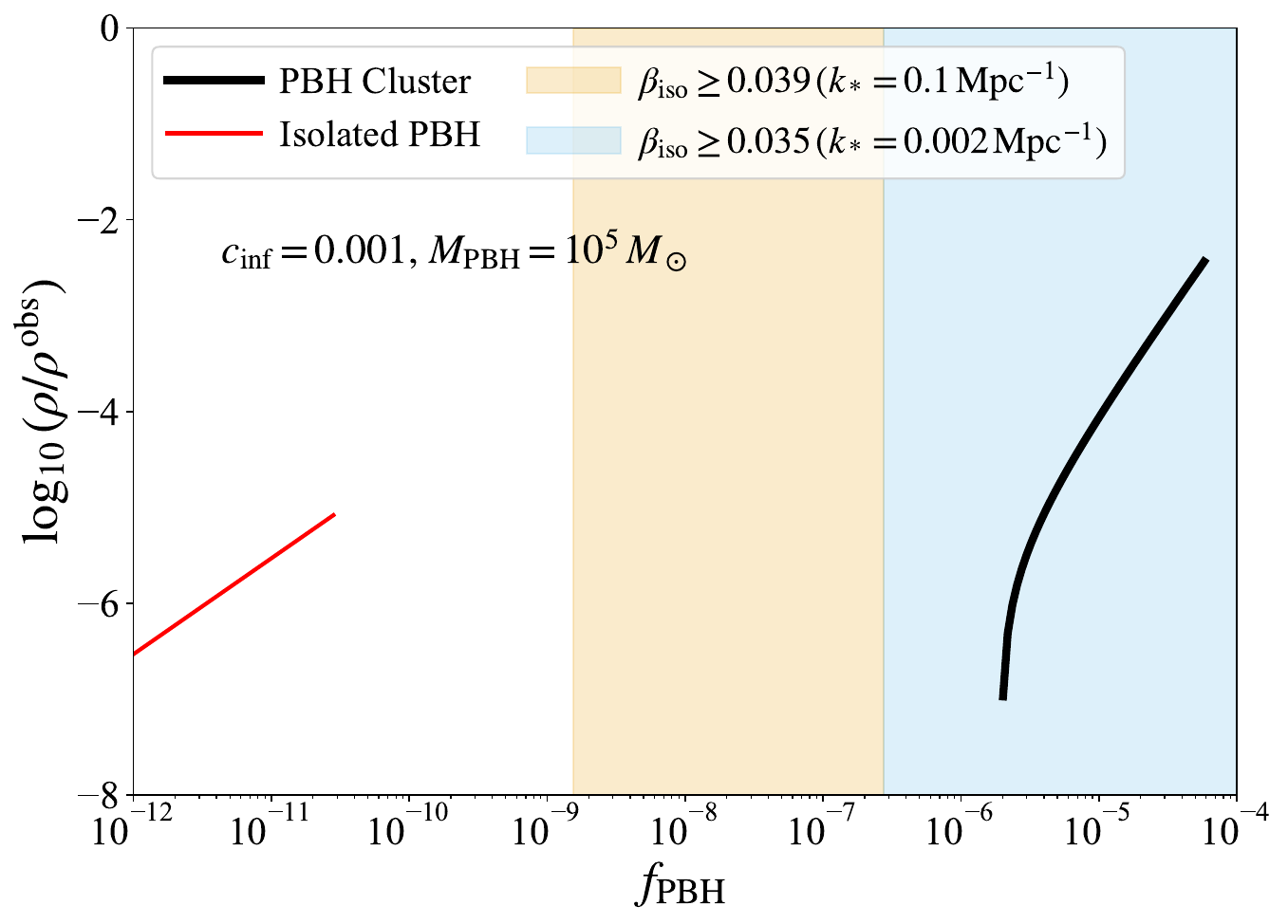}\quad
\includegraphics[width=0.48\textwidth]{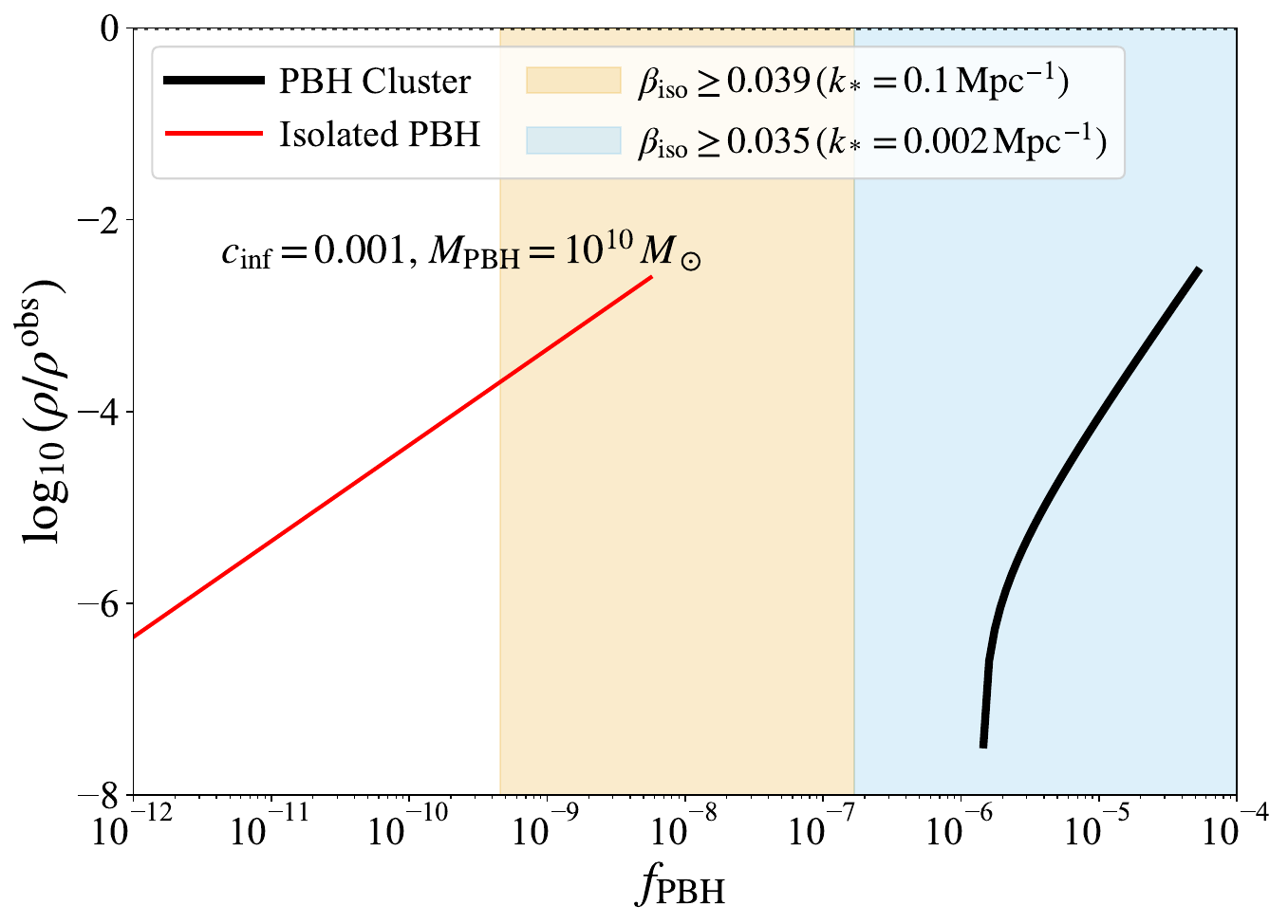}
\caption{The comoving cumulative stellar mass density $\rho/\rho^{\rm obs}$ arising from the seed effect as a function of the PBH abundance $f_\pb$, with the parameter $c_{\rm inf}=0.001$ and different PBH masses $M_\pb=10^{-10}$, $10^{-5}$, $10^5$, and $10^{10}~M_\odot$. The black and red lines represent the cases of compact PBH clusters and isolated PBHs, respectively. In all cases shown, the predicted stellar mass density remains below $\rho/\rho^{\rm obs}=1$. The blue shaded region is excluded by the CMB isocurvature constraint at the benchmark pivot scale $k_*=0.002\,{\rm Mpc}^{-1}$, while the orange shaded region indicates the corresponding excluded region for $k_*=0.1\,{\rm Mpc}^{-1}$, illustrating the dependence of the constraint on the choice of pivot scale. The star formation efficiency is fixed as $\epsilon_*=0.3$. For compact PBH clusters, significant enhancement of galaxy formation occurs only in the parameter regions constrained by the CMB isocurvature bounds for both $k_*=0.002$ and $0.1\,{\rm Mpc}^{-1}$. Moreover, because compact PBH clusters are rare, the number of galaxies formed through this localized seed mechanism is expected to be much smaller than the number of galaxies observed by the JWST. For isolated supermassive PBHs, those with $M_\pb\gtrsim10^4\,M_\odot$ can contribute to galaxy formation through the seed effect; however, owing to their extremely low abundance, they are insufficient to reproduce the stellar mass density inferred from the JWST observations. } 
\label{fig:rhoseedfPBH_00}
\end{figure}

Figure.~\ref{fig:rhoseedfPBH_00} shows the comoving cumulative stellar mass density $\rho/\rho^{\rm obs}$ generated by the PBH clusters (black line) and the isolated supermassive PBHs (red line), with $\ep_*=0.3$ and $c_{\rm inf}=0.001$. The isocurvature fraction constraint $\beta_{\rm iso}<0.035$ at $k_*=0.002\,{\rm Mpc}^{-1}$ is chosen as our benchmark, with the constraint $\beta_{\rm iso}<0.039$ at $k_*=0.1\,{\rm Mpc}^{-1}$ also considered. For the compact PBH clusters, we restrict our analysis to the nonlinear regime $k_{\rm cut}^{\rm cl}<k\leq k_{\max}$ and require $N_{\min}<{4\pi r_{\max}^3}\bar{n}_{\rm PBH}/{3}$. When $f_\pb$ becomes sufficiently large, the assumptions underlying the seed effect break down, resulting in the endpoint of the black lines in Fig.~\ref{fig:rhoseedfPBH_00}. The significant enhancement of galaxy formation occurs only in the parameter regions that are constrained by the CMB isocurvature bounds, for both $k_*=0.002$ and $0.1\,{\rm Mpc}^{-1}$. Moreover, because compact PBH clusters are rare, the corresponding number of galaxies formed through this localized seed mechanism is expected to be much smaller than the number of galaxies observed by the JWST. These results indicate that the seed effect of compact PBH clusters is insufficient to account for the JWST observations. For comparison, we also consider the limiting case of the isolated supermassive PBHs and restrict our calculation to the complementary large-scale regime with $k<k_{\rm cut}^{\rm cl}$. We find that, when $M_\pb\gtrsim10^4\,M_\odot$, an individual PBH can contribute to the formation of galaxies by the seed effect, and the corresponding $\rho/\rho^{\rm obs}$ approximately follows a power-law dependence on $f_{\rm PBH}$. Nevertheless, due to their extremely low abundance, isolated supermassive PBHs are insufficient to reproduce the stellar mass density inferred from the JWST observations. 

\begin{figure}[htb]
\centering
\includegraphics[width=0.48\textwidth]{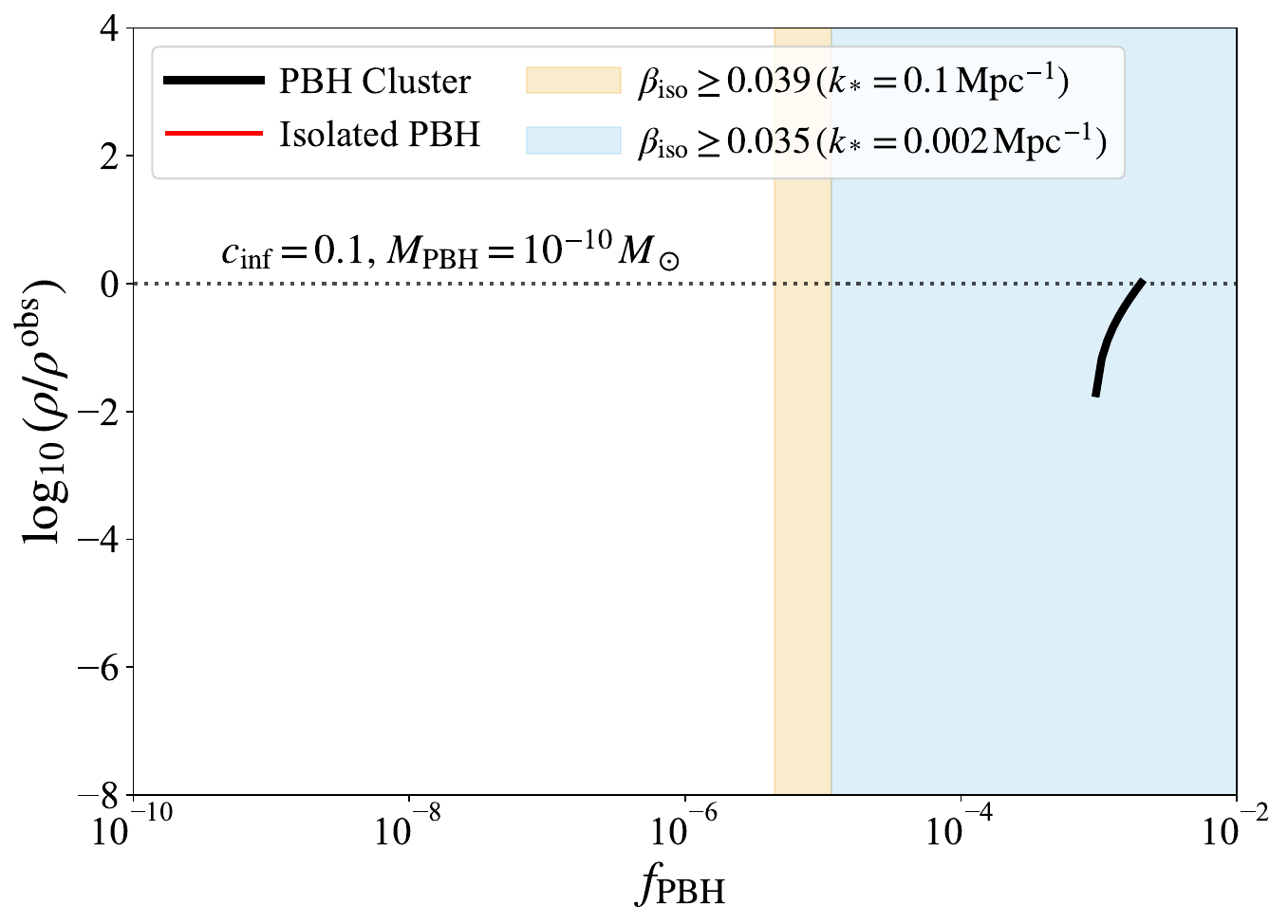}\quad
\includegraphics[width=0.48\textwidth]{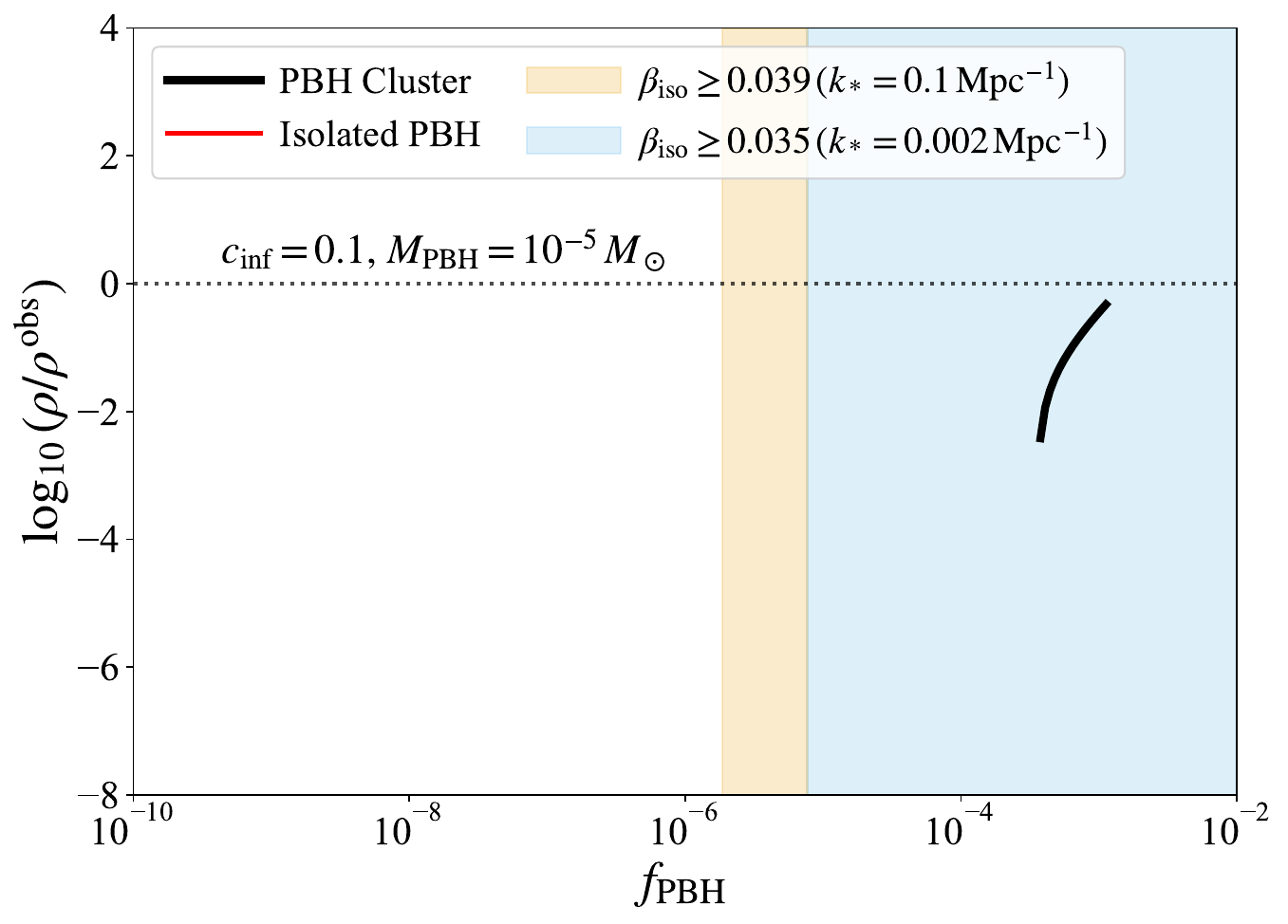}\\
\includegraphics[width=0.48\textwidth]{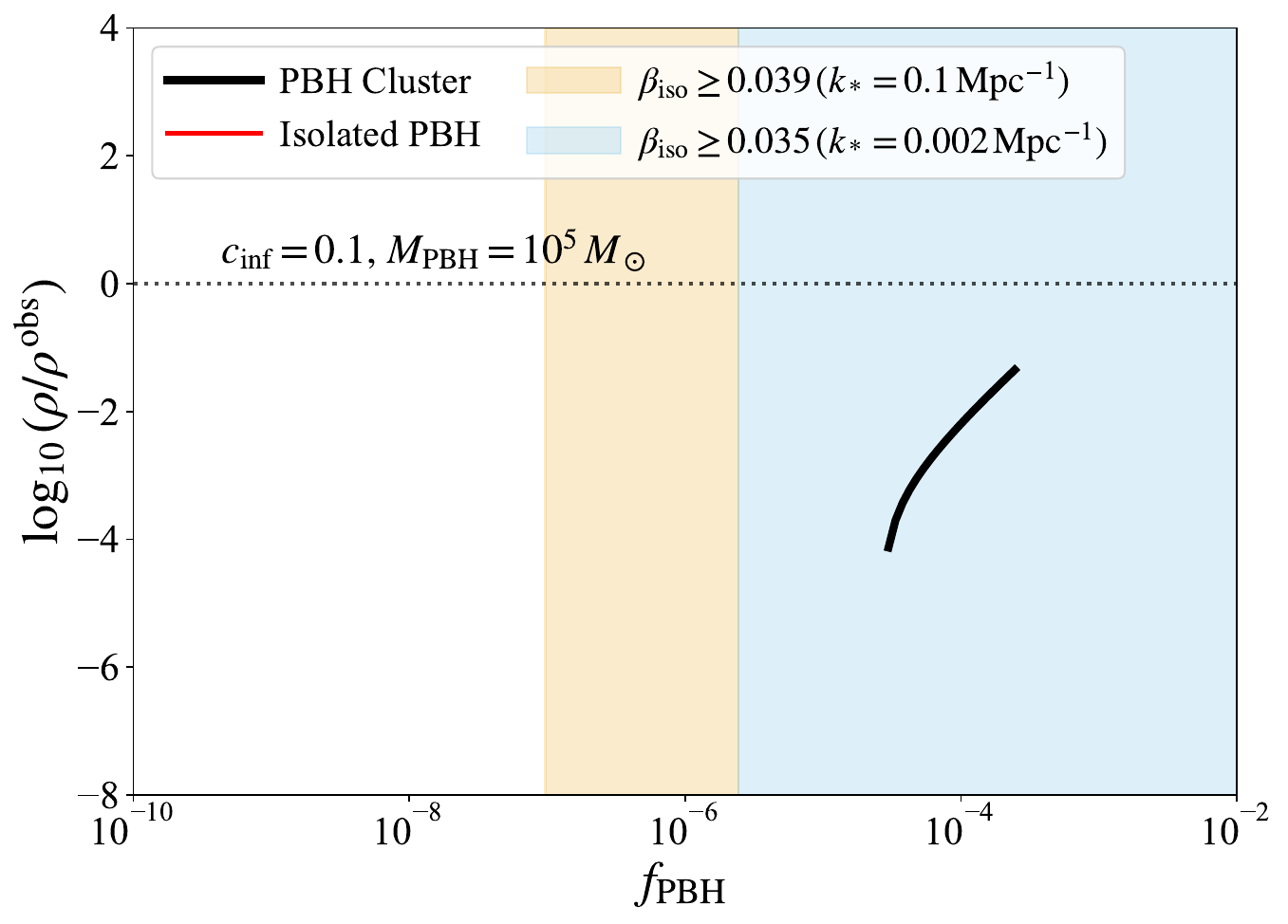}\quad
\includegraphics[width=0.48\textwidth]{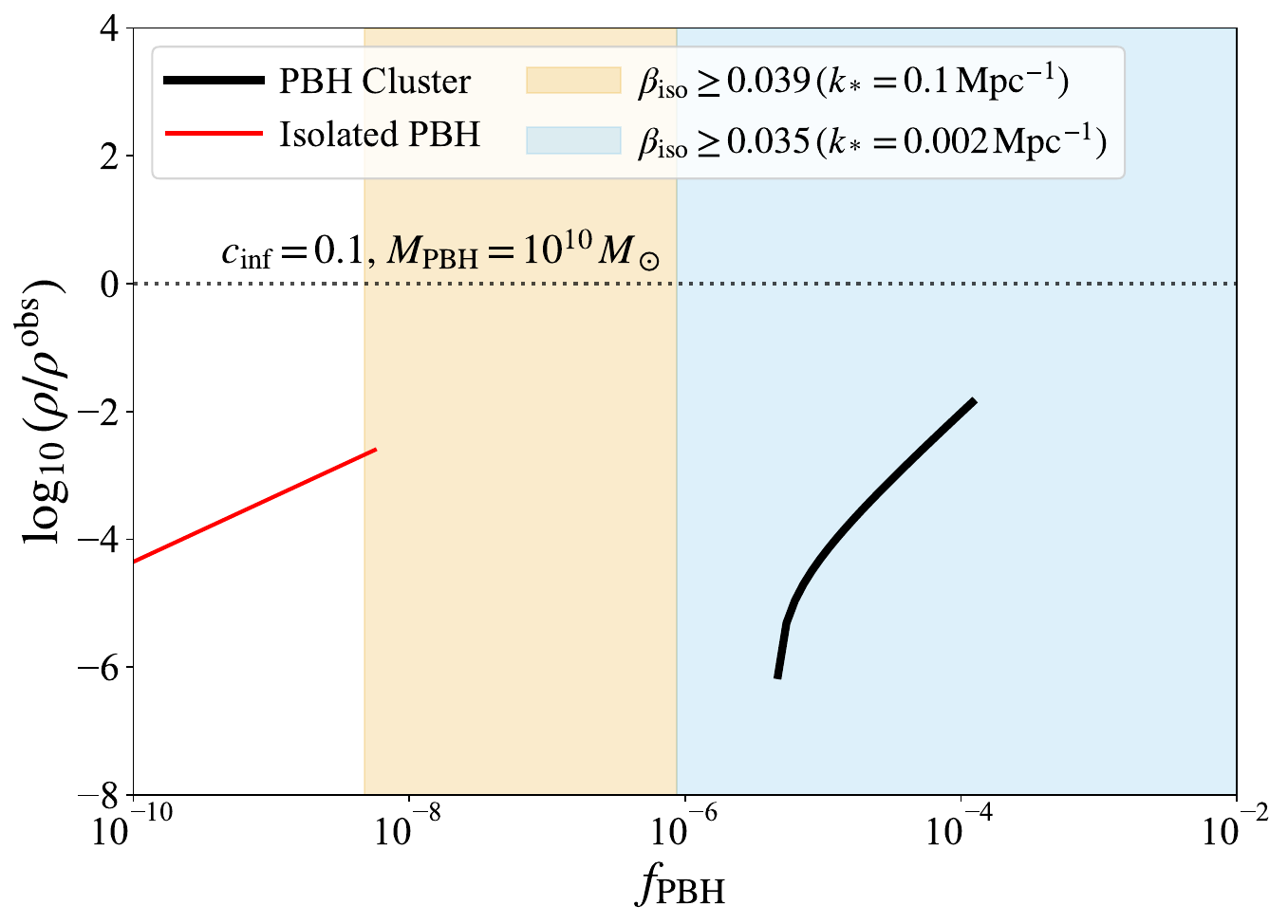}
\caption{Same as Fig.~\ref{fig:rhoseedfPBH_00}, but with the parameter $c_{\rm inf}=0.1$. For compact PBH clusters, the comoving cumulative stellar mass density $\rho/\rho^{\rm obs}$ increases as the PBH mass decreases and can reach 1 for $M_\pb\lesssim10^{-10}\,M_\odot$. However, achieving $\rho/\rho^{\rm obs}\simeq1$ requires $f_\pb\gtrsim10^{-3}$, which lies in the parameter regions excluded by the CMB isocurvature constraints at both $k_*=0.002$ and $0.1\,{\rm Mpc}^{-1}$. Therefore, the seed effect of compact PBH clusters remains insufficient to reproduce the stellar mass density inferred from the JWST observations. } 
\label{fig:rhoseedfPBH}
\end{figure}

Moreover, the case with $c_{\rm int}=0.1$ is illustrated in Fig.~\ref{fig:rhoseedfPBH}. As the PBH mass decreases, the comoving cumulative stellar mass density $\rho/\rho^{\rm obs}$ increases, and can reach 1 when $M_\pb\lesssim10^{-10}\,M_\odot$. However, this requires $f_\pb\gtrsim10^{-3}$, which lies in the parameter regions excluded by the CMB isocurvature constraints. Therefore, the seed effect remains insufficient to reproduce the JWST-inferred stellar mass density for $c_{\rm inf}=0.1$. 

\section{Conclusion} \label{sec:con}

In this work, we have investigated the impact of PBH clustering, generated by a light spectator field during inflation, on early cosmic structure formation to interpret the recent JWST observations of massive galaxy candidates at high redshifts $7\leq z\leq10$. The spectator field with a small dimensionless mass parameter $c_{\rm inf}\ll1$ can produce PBH clustering. 
By computing the PBH formation probability and the two-point correlation function $\xi_\pb$ within the spectator field framework, we have analyzed two potential channels, through which PBHs accelerate early structure formation: the enhancement of the matter power spectrum via clustered primordial isocurvature perturbations, and the localized seed effect caused by compact PBH clusters. Our main findings are summarized as follows.

On the one hand, owing to PBH clustering, the resulting primordial isocurvature perturbations on linear scales with $k_{\min}\leq k\leq k_{\rm cut}^{\rm cl}$ can significantly enhance the matter power spectrum during the matter-dominated era. By incorporating this enhancement into the Sheth--Tormen halo mass function, we find that the resulting stellar mass density is sensitive to the choice of the pivot scale used for the CMB isocurvature constraint. Adopting $\beta_{\rm iso}<0.035$ at the benchmark pivot scale $k_*=0.002\,{\rm Mpc}^{-1}$, we find that, the predicted comoving cumulative stellar mass density by PBH clustering at $z=8$ can reach the level inferred from the JWST observations, while satisfying the corresponding isocurvature constraint. In contrast, when the constraint at $k_*=0.1\,{\rm Mpc}^{-1}$ is imposed, the allowed enhancement from PBH clustering is strongly suppressed, demonstrating a pronounced dependence of the inferred parameter space on the choice of the pivot scale. This dependence is a result that the PBH-induced isocurvature power spectrum is scale-dependent. For comparison, the Poisson distributed (unclustered) PBHs require very large masses, $M_{\rm PBH}\geq1.79\times10^5\,M_\odot$, to reproduce the JWST-inferred stellar mass density. Such a parameter region is already strongly constrained by the CMB $\mu$-distortion, Lyman-$\alpha$ forest, dynamical friction, and X-ray binaries. Meanwhile, the standard $\Lambda$CDM model alone remains insufficient, yielding $\rho/\rho^{\rm obs}=4.78\times10^{-2}$.

On the other hand, on nonlinear small scales with $k_{\rm cut}^{\rm cl}<k\leq k_{\max}$, compact PBH clusters can act as overdense seeds and trigger localized matter accretion through their gravitational potential. Using the Press--Schechter formalism to evaluate the number variance $\sigma_{\rm cl}^2(\bar{N})$ of PBH clusters, we find that the individual compact PBH clusters, as well as the isolated supermassive PBHs with $M_{\rm PBH}\gtrsim10^4\,M_\odot$, can contribute to galaxy formation through the seed effect. However, compact PBH clusters are rare, and the parameter regions in which they can provide a significant enhancement are constrained by the CMB isocurvature bounds. Consequently, the number of galaxies formed through this localized seed mechanism is insufficient to account for the abundant high-redshift galaxies inferred from the JWST observations. Consequently, the seed effect only provides a subdominant contribution to early galaxy formation in the parameter space allowed by the current constraints.

Overall, our results show that PBH clustering induced by the spectator field can substantially enhance early structure formation through clustered primordial isocurvature perturbations. In particular, under the benchmark CMB isocurvature constraint at $k_*=0.002\,{\rm Mpc}^{-1}$, this mechanism can reproduce the stellar mass density inferred from the JWST observations with almost no impact of the PBH masses. However, the viability of this interpretation is sensitive to the pivot scale adopted for the CMB isocurvature constraint: the corresponding allowed enhancement is significantly reduced when the constraint at $k_*=0.1\,{\rm Mpc}^{-1}$ is applied. Therefore, PBH clustering provides a possible mechanism for accelerating early structure formation, but whether it can fully account for the JWST-inferred stellar mass density cannot be established independently of the treatment and the scale-dependence of the CMB isocurvature constraint. 

Future observations, including spectroscopic confirmation of high-redshift galaxy candidates and additional constraints from scalar-induced gravitational waves and 21~cm cosmology, will further test the spectator field PBH clustering scenario and help determine whether the enhanced early structure formation induced by PBH clustering can account for the JWST observations.

\acknowledgments

We deeply thank Xun Chen and Teruaki Suyama for very fruitful discussions. This work is supported by the National Key R\&D Program of China (Grant No. 2022YFF0503304), the National Natural Science Foundation of China (Grants Nos. 12373002, 12220101003, 11773075), the National Natural Science Foundation of Liaoning Province, China (Grant No. 2025-MS-034), and the Youth Innovation Promotion Association of Chinese Academy of Sciences (Grant No. 2016288). 


\bibliography{ref}

\providecommand{\href}[2]{#2}\begingroup\raggedright\begin{thebibliography}{10}

\bibitem{2022ApJ...940L..14N}
R.P.~{Naidu}, P.A.~{Oesch}, P.~{van Dokkum}, E.J.~{Nelson}, K.A.~{Suess},
  G.~{Brammer} et~al., \emph{{Two Remarkably Luminous Galaxy Candidates at
  $z\approx 10$--$12$ Revealed by JWST}},
  \href{https://doi.org/10.3847/2041-8213/ac9b22}{\emph{Astrophys. J. Lett.}
  {\bfseries 940} (2022) L14}
  [\href{https://arxiv.org/abs/arXiv:2207.09434}{{\ttfamily
  arXiv:2207.09434}}].

\bibitem{adams2023discovery}
N.~Adams, C.~Conselice, L.~Ferreira, D.~Austin, J.~Trussler,
  I.~Juod{\v{z}}balis et~al., \emph{Discovery and properties of ultra-high
  redshift galaxies ($9< z< 12$) in the jwst ero smacs 0723 field},
  \href{https://doi.org/10.1093/mnras/stac3347}{\emph{Mon. Not. Roy. Astron.
  Soc.} {\bfseries 518} (2023) 4755}
  [\href{https://arxiv.org/abs/arXiv:2207.11217}{{\ttfamily
  arXiv:2207.11217}}].

\bibitem{Yan:2022sxd}
H.~Yan, Z.~Ma, C.~Ling, C.~Cheng and J.-s.~Huang, \emph{{First Batch of
  $z\approx 11$--$20$ Candidate Objects Revealed by the James Webb Space
  Telescope Early Release Observations on SMACS 0723-73}},
  \href{https://doi.org/10.3847/2041-8213/aca80c}{\emph{Astrophys. J. Lett.}
  {\bfseries 942} (2023) L9}
  [\href{https://arxiv.org/abs/arXiv:2207.11558}{{\ttfamily
  arXiv:2207.11558}}].

\bibitem{2023MNRAS.519.1201A}
H.~{Atek}, M.~{Shuntov}, L.J.~{Furtak}, J.~{Richard}, J.-P.~{Kneib},
  G.~{Mahler} et~al., \emph{{Revealing galaxy candidates out to $z\sim16$ with
  JWST observations of the lensing cluster SMACS0723}},
  \href{https://doi.org/10.1093/mnras/stac3144}{\emph{Mon. Not. Roy. Astron.
  Soc.} {\bfseries 519} (2023) 1201}
  [\href{https://arxiv.org/abs/arXiv:2207.12338}{{\ttfamily
  arXiv:2207.12338}}].

\bibitem{2023MNRAS.518.6011D}
C.T.~{Donnan}, D.J.~{McLeod}, J.S.~{Dunlop}, R.J.~{McLure}, A.C.~{Carnall},
  R.~{Begley} et~al., \emph{{The evolution of the galaxy UV luminosity function
  at redshifts $z\backsimeq 8$--$15$ from deep JWST and ground-based
  near-infrared imaging}},
  \href{https://doi.org/10.1093/mnras/stac3472}{\emph{Mon. Not. Roy. Astron.
  Soc.} {\bfseries 518} (2023) 6011}
  [\href{https://arxiv.org/abs/arXiv:2207.12356}{{\ttfamily
  arXiv:2207.12356}}].

\bibitem{Labbe:2022ahb}
I.~Labb\'{e} et~al., \emph{{A population of red candidate massive galaxies
  {\textasciitilde}600 Myr after the Big Bang}},
  \href{https://doi.org/10.1038/s41586-023-05786-2}{\emph{Nature} {\bfseries
  616} (2023) 266} [\href{https://arxiv.org/abs/arXiv:2207.12446}{{\ttfamily
  arXiv:2207.12446}}].

\bibitem{2022ApJ...940L..55F}
S.L.~{Finkelstein}, M.B.~{Bagley}, P.A.~{Haro}, M.~{Dickinson},
  H.C.~{Ferguson}, J.S.~{Kartaltepe} et~al., \emph{{A Long Time Ago in a Galaxy
  Far, Far Away: A Candidate $z\sim12$ Galaxy in Early JWST CEERS Imaging}},
  \href{https://doi.org/10.3847/2041-8213/ac966e}{\emph{Astrophys. J. Lett.}
  {\bfseries 940} (2022) L55}
  [\href{https://arxiv.org/abs/arXiv:2207.12474}{{\ttfamily
  arXiv:2207.12474}}].

\bibitem{2023ApJS..265....5H}
Y.~{Harikane}, M.~{Ouchi}, M.~{Oguri}, Y.~{Ono}, K.~{Nakajima}, Y.~{Isobe}
  et~al., \emph{{A Comprehensive Study of Galaxies at $z\sim9$--$16$ Found in
  the Early JWST Data: Ultraviolet Luminosity Functions and Cosmic Star
  Formation History at the Pre-reionization Epoch}},
  \href{https://doi.org/10.3847/1538-4365/acaaa9}{\emph{Astrophys. J. Suppl. S}
  {\bfseries 265} (2023) 5}
  [\href{https://arxiv.org/abs/arXiv:2208.01612}{{\ttfamily
  arXiv:2208.01612}}].

\bibitem{Robertson:2022gdk}
B.E.~Robertson et~al., \emph{{Identification and properties of intense
  star-forming galaxies at redshifts z{\,}{\ensuremath{>}}{\,}10}},
  \href{https://doi.org/10.1038/s41550-023-01921-1}{\emph{Nature Astron.}
  {\bfseries 7} (2023) 611}
  [\href{https://arxiv.org/abs/arXiv:2212.04480}{{\ttfamily
  arXiv:2212.04480}}].

\bibitem{Robertson:2023nyv}
B.~Robertson et~al., \emph{{Earliest Galaxies in the JADES Origins Field:
  Luminosity Function and Cosmic Star Formation Rate Density 300 Myr after the
  Big Bang}}, \href{https://doi.org/10.3847/1538-4357/ad463d}{\emph{Astrophys.
  J.} {\bfseries 970} (2024) 31}
  [\href{https://arxiv.org/abs/arXiv:2312.10033}{{\ttfamily
  arXiv:2312.10033}}].

\bibitem{Perez-Gonzalez:2023wta}
P.G.~P{\'e}rez-Gonz{\'a}lez et~al., \emph{{Life beyond 30: Probing the
  {\ensuremath{-}}20 {\ensuremath{<}} M $_{UV}$ {\ensuremath{<}}
  {\ensuremath{-}}17 Luminosity Function at 8 {\ensuremath{<}} z
  {\ensuremath{<}} 13 with the NIRCam Parallel Field of the MIRI Deep Survey}},
  \href{https://doi.org/10.3847/2041-8213/acd9d0}{\emph{Astrophys. J. Lett.}
  {\bfseries 951} (2023) L1}
  [\href{https://arxiv.org/abs/arXiv:2302.02429}{{\ttfamily
  arXiv:2302.02429}}].

\bibitem{2023NatAs...7..611R}
B.E.~{Robertson}, S.~{Tacchella}, B.D.~{Johnson}, K.~{Hainline}, L.~{Whitler},
  D.J.~{Eisenstein} et~al., \emph{{Identification and properties of intense
  star-forming galaxies at redshifts z > 10}},
  \href{https://doi.org/10.1038/s41550-023-01921-1}{\emph{Nature Astron.}
  {\bfseries 7} (2023) 611}
  [\href{https://arxiv.org/abs/arXiv:2212.04480}{{\ttfamily
  arXiv:2212.04480}}].

\bibitem{2023NatAs...7..622C}
E.~{Curtis-Lake}, S.~{Carniani}, A.~{Cameron}, S.~{Charlot}, P.~{Jakobsen},
  R.~{Maiolino} et~al., \emph{{Spectroscopic confirmation of four metal-poor
  galaxies at z = 10.3-13.2}},
  \href{https://doi.org/10.1038/s41550-023-01918-w}{\emph{Nature Astron.}
  {\bfseries 7} (2023) 622}
  [\href{https://arxiv.org/abs/arXiv:2212.04568}{{\ttfamily
  arXiv:2212.04568}}].

\bibitem{2023ApJ...951L..22A}
P.~{Arrabal Haro}, M.~{Dickinson}, S.L.~{Finkelstein}, S.~{Fujimoto},
  V.~{Fern{\'a}ndez}, J.S.~{Kartaltepe} et~al., \emph{{Spectroscopic
  Confirmation of CEERS NIRCam-selected Galaxies at $z \simeq 8$--10}},
  \href{https://doi.org/10.3847/2041-8213/acdd54}{\emph{Astrophys. J. Lett.}
  {\bfseries 951} (2023) L22}
  [\href{https://arxiv.org/abs/arXiv:2304.05378}{{\ttfamily
  arXiv:2304.05378}}].

\bibitem{2024Natur.633..318C}
S.~{Carniani}, K.~{Hainline}, F.~{D'Eugenio}, D.J.~{Eisenstein}, P.~{Jakobsen},
  J.~{Witstok} et~al., \emph{{Spectroscopic confirmation of two luminous
  galaxies at a redshift of 14}},
  \href{https://doi.org/10.1038/s41586-024-07860-9}{\emph{Nature} {\bfseries
  633} (2024) 318} [\href{https://arxiv.org/abs/arXiv:2405.18485}{{\ttfamily
  arXiv:2405.18485}}].

\bibitem{2013ApJ...770...57B}
P.S.~{Behroozi}, R.H.~{Wechsler} and C.~{Conroy}, \emph{{The Average Star
  Formation Histories of Galaxies in Dark Matter Halos from $z = 0$--$8$}},
  \href{https://doi.org/10.1088/0004-637X/770/1/57}{\emph{Astrophys. J.}
  {\bfseries 770} (2013) 57}
  [\href{https://arxiv.org/abs/arXiv:1207.6105}{{\ttfamily arXiv:1207.6105}}].

\bibitem{2019MNRAS.488.3143B}
P.~{Behroozi}, R.H.~{Wechsler}, A.P.~{Hearin} and C.~{Conroy},
  \emph{{UNIVERSEMACHINE: The correlation between galaxy growth and dark matter
  halo assembly from $z = 0$--$10$}},
  \href{https://doi.org/10.1093/mnras/stz1182}{\emph{Mon. Not. Roy. Astron.
  Soc.} {\bfseries 488} (2019) 3143}
  [\href{https://arxiv.org/abs/arXiv:1806.07893}{{\ttfamily
  arXiv:1806.07893}}].

\bibitem{2022ApJ...938L..10I}
K.~{Inayoshi}, Y.~{Harikane}, A.K.~{Inoue}, W.~{Li} and L.C.~{Ho}, \emph{{A
  Lower Bound of Star Formation Activity in Ultra-high-redshift Galaxies
  Detected with JWST: Implications for Stellar Populations and Radiation
  Sources}}, \href{https://doi.org/10.3847/2041-8213/ac9310}{\emph{Astrophys.
  J. Lett.} {\bfseries 938} (2022) L10}
  [\href{https://arxiv.org/abs/arXiv:2208.06872}{{\ttfamily
  arXiv:2208.06872}}].

\bibitem{Lovell:2022bhx}
C.C.~Lovell, I.~Harrison, Y.~Harikane, S.~Tacchella and S.M.~Wilkins,
  \emph{{Extreme value statistics of the halo and stellar mass distributions at
  high redshift: are JWST results in tension with $\Lambda$CDM?}},
  \href{https://doi.org/10.1093/mnras/stac3224}{\emph{Mon. Not. Roy. Astron.
  Soc.} {\bfseries 518} (2022) 2511}
  [\href{https://arxiv.org/abs/arXiv:2208.10479}{{\ttfamily
  arXiv:2208.10479}}].

\bibitem{Forconi:2023izg}
M.~Forconi, Ruchika, A.~Melchiorri, O.~Mena and N.~Menci, \emph{{Do the early
  galaxies observed by JWST disagree with Planck's CMB polarization
  measurements?}},
  \href{https://doi.org/10.1088/1475-7516/2023/10/012}{\emph{J. Cosmology
  Astropart. Phys.} {\bfseries 10} (2023) 012}
  [\href{https://arxiv.org/abs/arXiv:2306.07781}{{\ttfamily
  arXiv:2306.07781}}].

\bibitem{Menci:2022wia}
N.~Menci, M.~Castellano, P.~Santini, E.~Merlin, A.~Fontana and F.~Shankar,
  \emph{{High-redshift Galaxies from Early JWST Observations: Constraints on
  Dark Energy Models}},
  \href{https://doi.org/10.3847/2041-8213/ac96e9}{\emph{Astrophys. J. Lett.}
  {\bfseries 938} (2022) L5}
  [\href{https://arxiv.org/abs/arXiv:2208.11471}{{\ttfamily
  arXiv:2208.11471}}].

\bibitem{Gong:2022qjx}
Y.~Gong, B.~Yue, Y.~Cao and X.~Chen, \emph{{Fuzzy Dark Matter as a Solution to
  Reconcile the Stellar Mass Density of High-$z$ Massive Galaxies and
  Reionization History}},
  \href{https://doi.org/10.3847/1538-4357/acc109}{\emph{Astrophys. J.}
  {\bfseries 947} (2023) 28}
  [\href{https://arxiv.org/abs/arXiv:2209.13757}{{\ttfamily
  arXiv:2209.13757}}].

\bibitem{Biagetti:2022ode}
M.~Biagetti, G.~Franciolini and A.~Riotto, \emph{{High-redshift JWST
  Observations and Primordial Non-Gaussianity}},
  \href{https://doi.org/10.3847/1538-4357/acb5ea}{\emph{Astrophys. J.}
  {\bfseries 944} (2023) 113}
  [\href{https://arxiv.org/abs/arXiv:2210.04812}{{\ttfamily
  arXiv:2210.04812}}].

\bibitem{Wang:2022jvx}
D.~Wang and Y.~Liu, \emph{{JWST high redshift galaxy observations have a strong
  tension with Planck CMB measurements}},
  \href{https://arxiv.org/abs/arXiv:2301.00347}{{\ttfamily arXiv:2301.00347}}.

\bibitem{Parashari:2023cui}
P.~Parashari and R.~Laha, \emph{{Primordial power spectrum in light of JWST
  observations of high redshift galaxies}},
  \href{https://doi.org/10.1093/mnrasl/slad107}{\emph{Mon. Not. Roy. Astron.
  Soc.} {\bfseries 526} (2023) L63}
  [\href{https://arxiv.org/abs/arXiv:2305.00999}{{\ttfamily
  arXiv:2305.00999}}].

\bibitem{Dayal:2023nwi}
P.~Dayal and S.K.~Giri, \emph{{Warm dark matter constraints from the JWST}},
  \href{https://doi.org/10.1093/mnras/stae176}{\emph{Mon. Not. Roy. Astron.
  Soc.} {\bfseries 528} (2024) 2784}
  [\href{https://arxiv.org/abs/arXiv:2303.14239}{{\ttfamily
  arXiv:2303.14239}}].

\bibitem{2023MNRAS.526.2542C}
Y.~{Chen}, H.J.~{Mo} and K.~{Wang}, \emph{{Massive dark matter haloes at high
  redshift: implications for observations in the JWST era}},
  \href{https://doi.org/10.1093/mnras/stad2866}{\emph{Mon. Not. Roy. Astron.
  Soc.} {\bfseries 526} (2023) 2542}
  [\href{https://arxiv.org/abs/arXiv:2304.13890}{{\ttfamily
  arXiv:2304.13890}}].

\bibitem{Lopez-Corredoira:2024pgl}
M.~Lopez-Corredoira, F.~Melia, J.J.~Wei and C.Y.~Gao, \emph{{Age of Massive
  Galaxies at Redshift 8}},
  \href{https://doi.org/10.3847/1538-4357/ad4f86}{\emph{Astrophys. J.}
  {\bfseries 970} (2024) 63}
  [\href{https://arxiv.org/abs/arXiv:2405.12665}{{\ttfamily
  arXiv:2405.12665}}].

\bibitem{2026arXiv260418683P}
F.~{Prada}, P.~{Behroozi}, T.~{Ishiyama}, E.~{P{\'e}rez}, A.~{Klypin},
  X.~{Wang} et~al., \emph{{Galaxies at z > 10: {\ensuremath{\Lambda}}CDM
  predicts increased Star Formation Efficiency}},
  \href{https://arxiv.org/abs/arXiv:2604.18683}{{\ttfamily arXiv:2604.18683}}.

\bibitem{Ferrara:2022dqw}
A.~Ferrara, A.~Pallottini and P.~Dayal, \emph{{On the stunning abundance of
  super-early, luminous galaxies revealed by JWST}},
  \href{https://doi.org/10.1093/mnras/stad1095}{\emph{Mon. Not. Roy. Astron.
  Soc.} {\bfseries 522} (2023) 3986}
  [\href{https://arxiv.org/abs/arXiv:2208.00720}{{\ttfamily
  arXiv:2208.00720}}].

\bibitem{Ziparo:2022rir}
F.~Ziparo, A.~Ferrara, L.~Sommovigo and M.~Kohandel, \emph{{Blue monsters. Why
  are JWST super-early, massive galaxies so blue?}},
  \href{https://doi.org/10.1093/mnras/stad125}{\emph{Mon. Not. Roy. Astron.
  Soc.} {\bfseries 520} (2023) 2445}
  [\href{https://arxiv.org/abs/arXiv:2209.06840}{{\ttfamily
  arXiv:2209.06840}}].

\bibitem{2025A&A...702A..33M}
V.~{Markov}, S.~{Gallerani}, A.~{Pallottini}, M.~{Brada{\v{c}}}, S.~{Carniani},
  R.~{Tripodi} et~al., \emph{{Unveiling the trends between dust attenuation and
  galaxy properties at z {\ensuremath{\sim}} 2{\ensuremath{-}}12 with the James
  Webb Space Telescope}},
  \href{https://doi.org/10.1051/0004-6361/202555182}{\emph{Astron. Astrophys.}
  {\bfseries 702} (2025) A33}
  [\href{https://arxiv.org/abs/arXiv:2504.12378}{{\ttfamily
  arXiv:2504.12378}}].

\bibitem{2023A&A...679A..12M}
V.~{Markov}, S.~{Gallerani}, A.~{Pallottini}, L.~{Sommovigo}, S.~{Carniani},
  A.~{Ferrara} et~al., \emph{{Dust attenuation law in JWST galaxies at z
  {\ensuremath{\sim}} 7-8}},
  \href{https://doi.org/10.1051/0004-6361/202346723}{\emph{Astron. Astrophys.}
  {\bfseries 679} (2023) A12}
  [\href{https://arxiv.org/abs/arXiv:2304.11178}{{\ttfamily
  arXiv:2304.11178}}].

\bibitem{2025NatAs...9..458M}
V.~{Markov}, S.~{Gallerani}, A.~{Ferrara}, A.~{Pallottini}, E.~{Parlanti},
  F.D.~{Mascia} et~al., \emph{{The evolution of dust attenuation in z
  {\ensuremath{\approx}} 2-12 galaxies observed by JWST}},
  \href{https://doi.org/10.1038/s41550-024-02426-1}{\emph{Nature Astron.}
  {\bfseries 9} (2025) 458}
  [\href{https://arxiv.org/abs/arXiv:2402.05996}{{\ttfamily
  arXiv:2402.05996}}].

\bibitem{2023Natur.622..707A}
P.~{Arrabal Haro}, M.~{Dickinson}, S.L.~{Finkelstein}, J.S.~{Kartaltepe},
  C.T.~{Donnan}, D.~{Burgarella} et~al., \emph{{Confirmation and refutation of
  very luminous galaxies in the early Universe}},
  \href{https://doi.org/10.1038/s41586-023-06521-7}{\emph{Nature} {\bfseries
  622} (2023) 707} [\href{https://arxiv.org/abs/arXiv:2303.15431}{{\ttfamily
  arXiv:2303.15431}}].

\bibitem{Boylan-Kolchin:2022kae}
M.~Boylan-Kolchin, \emph{{Stress testing {\ensuremath{\Lambda}}CDM with
  high-redshift galaxy candidates}},
  \href{https://doi.org/10.1038/s41550-023-01937-7}{\emph{Nature Astron.}
  {\bfseries 7} (2023) 731}
  [\href{https://arxiv.org/abs/arXiv:2208.01611}{{\ttfamily
  arXiv:2208.01611}}].

\bibitem{2023MNRAS.519..843M}
J.~{Mirocha} and S.R.~{Furlanetto}, \emph{{Balancing the efficiency and
  stochasticity of star formation with dust extinction in $z\gtrsim 10$
  galaxies observed by JWST}},
  \href{https://doi.org/10.1093/mnras/stac3578}{\emph{Mon. Not. Roy. Astron.
  Soc.} {\bfseries 519} (2023) 843}
  [\href{https://arxiv.org/abs/arXiv:2208.12826}{{\ttfamily
  arXiv:2208.12826}}].

\bibitem{Wang:2023ros}
P.~Wang, B.-Y.~Su, L.~Zu, Y.~Yang and L.~Feng, \emph{{Exploring the dark energy
  equation of state with JWST}},
  \href{https://doi.org/10.1140/epjp/s13360-024-05276-y}{\emph{Eur. Phys. J.
  Plus} {\bfseries 139} (2024) 711}
  [\href{https://arxiv.org/abs/arXiv:2307.11374}{{\ttfamily
  arXiv:2307.11374}}].

\bibitem{Dekel:2023ddd}
A.~Dekel, K.C.~Sarkar, Y.~Birnboim, N.~Mandelker and Z.~Li, \emph{{Efficient
  formation of massive galaxies at cosmic dawn by feedback-free starbursts}},
  \href{https://doi.org/10.1093/mnras/stad1557}{\emph{Mon. Not. Roy. Astron.
  Soc.} {\bfseries 523} (2023) 3201}
  [\href{https://arxiv.org/abs/arXiv:2303.04827}{{\ttfamily
  arXiv:2303.04827}}].

\bibitem{2023ApJ...955L..35S}
G.~{Sun}, C.-A.~{Faucher-Gigu{\`e}re}, C.C.~{Hayward}, X.~{Shen}, A.~{Wetzel}
  and R.K.~{Cochrane}, \emph{{Bursty Star Formation Naturally Explains the
  Abundance of Bright Galaxies at Cosmic Dawn}},
  \href{https://doi.org/10.3847/2041-8213/acf85a}{\emph{Astrophys. J. Lett.}
  {\bfseries 955} (2023) L35}
  [\href{https://arxiv.org/abs/arXiv:2307.15305}{{\ttfamily
  arXiv:2307.15305}}].

\bibitem{Zeldovich:1967lct}
Y.B.~Zel'dovich and I.D.~Novikov, \emph{{The Hypothesis of Cores Retarded
  during Expansion and the Hot Cosmological Model}}, {\emph{Soviet Astron. AJ
  (Engl. Transl.),} {\bfseries 10} (1967) 602}.

\bibitem{Hawking:1971ei}
S.~Hawking, \emph{{Gravitationally collapsed objects of very low mass}},
  \href{https://doi.org/10.1093/mnras/152.1.75}{\emph{Mon. Not. Roy. Astron.
  Soc.} {\bfseries 152} (1971) 75}.

\bibitem{Kohri:2014lza}
K.~Kohri, T.~Nakama and T.~Suyama, \emph{{Testing scenarios of primordial black
  holes being the seeds of supermassive black holes by ultracompact minihalos
  and CMB $\mu$-distortions}},
  \href{https://doi.org/10.1103/PhysRevD.90.083514}{\emph{Phys. Rev. D}
  {\bfseries 90} (2014) 083514}
  [\href{https://arxiv.org/abs/arXiv:1405.5999}{{\ttfamily arXiv:1405.5999}}].

\bibitem{Liu:2022bvr}
B.~Liu and V.~Bromm, \emph{{Accelerating Early Massive Galaxy Formation with
  Primordial Black Holes}},
  \href{https://doi.org/10.3847/2041-8213/ac927f}{\emph{Astrophys. J. Lett.}
  {\bfseries 937} (2022) L30}
  [\href{https://arxiv.org/abs/arXiv:2208.13178}{{\ttfamily
  arXiv:2208.13178}}].

\bibitem{Yuan:2023bvh}
G.-W.~Yuan, L.~Lei, Y.-Z.~Wang, B.~Wang, Y.-Y.~Wang, C.~Chen et~al.,
  \emph{{Rapidly growing primordial black holes as seeds of the massive
  high-redshift JWST Galaxies}},
  \href{https://arxiv.org/abs/arXiv:2303.09391}{{\ttfamily arXiv:2303.09391}}.

\bibitem{Huang:2024aog}
H.-L.~Huang, J.-Q.~Jiang and Y.-S.~Piao, \emph{{High-redshift JWST massive
  galaxies and the initial clustering of supermassive primordial black holes}},
  \href{https://doi.org/10.1103/PhysRevD.110.103540}{\emph{Phys. Rev. D}
  {\bfseries 110} (2024) 103540}
  [\href{https://arxiv.org/abs/arXiv:2407.15781}{{\ttfamily
  arXiv:2407.15781}}].

\bibitem{Hooper:2023nnl}
D.~Hooper, A.~Ireland, G.~Krnjaic and A.~Stebbins, \emph{{Supermassive
  primordial black holes from inflation}},
  \href{https://doi.org/10.1088/1475-7516/2024/04/021}{\emph{J. Cosmology
  Astropart. Phys.} {\bfseries 04} (2024) 021}
  [\href{https://arxiv.org/abs/arXiv:2308.00756}{{\ttfamily
  arXiv:2308.00756}}].

\bibitem{Dave:2026gsd}
S.~Dave, S.-F.~Yan, A.~Ilyas and Y.-F.~Cai, \emph{{Evading the CMB
  $\mu$-distortion bound on Supermassive Primordial Black Hole seeds with
  Non-Gaussian tails}},
  \href{https://arxiv.org/abs/arXiv:2607.03138}{{\ttfamily arXiv:2607.03138}}.

\bibitem{Meszaros:1975ef}
P.~M\'{e}sz\'{a}ros, \emph{{Primeval black holes and galaxy formation}},
  {\emph{Astron. Astrophys.} {\bfseries 38} (1975) 5}.

\bibitem{1983ApJ...268....1C}
B.J.~{Carr} and J.~{Silk}, \emph{{Can graininess in the early universe make
  galaxies?}}, \href{https://doi.org/10.1086/160924}{\emph{Astrophys. J.}
  {\bfseries 268} (1983) 1}.

\bibitem{Carr:1984ikr}
B.J.~Carr and M.J.~Rees., \emph{{Can pregalactic objects generate galaxies?}},
  \href{https://doi.org/10.1093/mnras/206.4.801}{\emph{Mon. Not. Roy. Astron.
  Soc.} {\bfseries 206} (1984) 801}.

\bibitem{Carr:2018rid}
B.~Carr and J.~Silk, \emph{{Primordial Black Holes as Generators of Cosmic
  Structures}}, \href{https://doi.org/10.1093/mnras/sty1204}{\emph{Mon. Not.
  Roy. Astron. Soc.} {\bfseries 478} (2018) 3756}
  [\href{https://arxiv.org/abs/arXiv:1801.00672}{{\ttfamily
  arXiv:1801.00672}}].

\bibitem{Hutsi:2022fzw}
G.~H\"utsi, M.~Raidal, J.~Urrutia, V.~Vaskonen and H.~Veerm\"ae, \emph{{Did
  JWST observe imprints of axion miniclusters or primordial black holes?}},
  \href{https://doi.org/10.1103/PhysRevD.107.043502}{\emph{Phys. Rev. D}
  {\bfseries 107} (2023) 043502}
  [\href{https://arxiv.org/abs/arXiv:2211.02651}{{\ttfamily
  arXiv:2211.02651}}].

\bibitem{Kannike:2017bxn}
K.~Kannike, L.~Marzola, M.~Raidal and H.~Veerm\"ae, \emph{{Single Field Double
  Inflation and Primordial Black Holes}},
  \href{https://doi.org/10.1088/1475-7516/2017/09/020}{\emph{J. Cosmol.
  Astropart. Phys.} {\bfseries 09} (2017) 020}
  [\href{https://arxiv.org/abs/arXiv:1705.06225}{{\ttfamily
  arXiv:1705.06225}}].

\bibitem{Ballesteros:2017fsr}
G.~Ballesteros and M.~Taoso, \emph{{Primordial black hole dark matter from
  single field inflation}},
  \href{https://doi.org/10.1103/PhysRevD.97.023501}{\emph{Phys. Rev. D}
  {\bfseries 97} (2018) 023501}
  [\href{https://arxiv.org/abs/arXiv:1709.05565}{{\ttfamily
  arXiv:1709.05565}}].

\bibitem{Gangopadhyay:2021kmf}
M.R.~Gangopadhyay, J.C.~Jain, D.~Sharma and Yogesh, \emph{{Production of
  primordial black holes via single field inflation and observational
  constraints}},
  \href{https://doi.org/10.1140/epjc/s10052-022-10796-x}{\emph{Eur. Phys. J. C}
  {\bfseries 82} (2022) 849}
  [\href{https://arxiv.org/abs/arXiv:2108.13839}{{\ttfamily
  arXiv:2108.13839}}].

\bibitem{Wang:2021kbh}
Q.~Wang, Y.-C.~Liu, B.-Y.~Su and N.~Li, \emph{{Primordial black holes from the
  perturbations in the inflaton potential in peak theory}},
  \href{https://doi.org/10.1103/PhysRevD.104.083546}{\emph{Phys. Rev. D}
  {\bfseries 104} (2021) 083546}
  [\href{https://arxiv.org/abs/arXiv:2111.10028}{{\ttfamily
  arXiv:2111.10028}}].

\bibitem{Liu:2021qky}
Y.-C.~Liu, Q.~Wang, B.-Y.~Su and N.~Li, \emph{{Primordial black holes from the
  perturbations in the inflaton potential}},
  \href{https://doi.org/10.1016/j.dark.2021.100905}{\emph{Phys. Dark Univ.}
  {\bfseries 34} (2021) 100905}.

\bibitem{Planck:2018vyg}
{\scshape Planck} collaboration, \emph{{Planck 2018 results. VI. Cosmological
  parameters}},
  \href{https://doi.org/10.1051/0004-6361/201833910}{\emph{Astron. Astrophys.}
  {\bfseries 641} (2020) A6}
  [\href{https://arxiv.org/abs/arXiv:1807.06209}{{\ttfamily
  arXiv:1807.06209}}].

\bibitem{Planck:2018jri}
{\scshape Planck} collaboration, \emph{{Planck 2018 results. X. Constraints on
  inflation}}, \href{https://doi.org/10.1051/0004-6361/201833887}{\emph{Astron.
  Astrophys.} {\bfseries 641} (2020) A10}
  [\href{https://arxiv.org/abs/arXiv:1807.06211}{{\ttfamily
  arXiv:1807.06211}}].

\bibitem{Carr:2020gox}
B.~Carr, K.~Kohri, Y.~Sendouda and J.~Yokoyama, \emph{{Constraints on
  primordial black holes}},
  \href{https://doi.org/10.1088/1361-6633/ac1e31}{\emph{Rept. Prog. Phys.}
  {\bfseries 84} (2021) 116902}
  [\href{https://arxiv.org/abs/arXiv:2002.12778}{{\ttfamily
  arXiv:2002.12778}}].

\bibitem{Murgia:2019duy}
R.~Murgia, G.~Scelfo, M.~Viel and A.~Raccanelli, \emph{{Lyman-$\alpha$ Forest
  Constraints on Primordial Black Holes as Dark Matter}},
  \href{https://doi.org/10.1103/PhysRevLett.123.071102}{\emph{Phys. Rev. Lett.}
  {\bfseries 123} (2019) 071102}
  [\href{https://arxiv.org/abs/arXiv:1903.10509}{{\ttfamily
  arXiv:1903.10509}}].

\bibitem{1999ApJ...516..195C}
B.J.~{Carr} and M.~{Sakellariadou}, \emph{{Dynamical Constraints on Dark Matter
  in Compact Objects}}, \href{https://doi.org/10.1086/307071}{\emph{Astrophys.
  J.} {\bfseries 516} (1999) 195}.

\bibitem{Inoue:2017csr}
Y.~Inoue and A.~Kusenko, \emph{{New X-ray bound on density of primordial black
  holes}}, \href{https://doi.org/10.1088/1475-7516/2017/10/034}{\emph{J.
  Cosmology Astropart. Phys.} {\bfseries 10} (2017) 034}
  [\href{https://arxiv.org/abs/arXiv:1705.00791}{{\ttfamily
  arXiv:1705.00791}}].

\bibitem{Nakama:2016kfq}
T.~Nakama, T.~Suyama and J.~Yokoyama, \emph{{Supermassive black holes formed by
  direct collapse of inflationary perturbations}},
  \href{https://doi.org/10.1103/PhysRevD.94.103522}{\emph{Phys. Rev. D}
  {\bfseries 94} (2016) 103522}
  [\href{https://arxiv.org/abs/arXiv:1609.02245}{{\ttfamily
  arXiv:1609.02245}}].

\bibitem{Shinohara:2021psq}
T.~Shinohara, T.~Suyama and T.~Takahashi, \emph{{Angular correlation as a novel
  probe of supermassive primordial black holes}},
  \href{https://doi.org/10.1103/PhysRevD.104.023526}{\emph{Phys. Rev. D}
  {\bfseries 104} (2021) 023526}
  [\href{https://arxiv.org/abs/arXiv:2103.13692}{{\ttfamily
  arXiv:2103.13692}}].

\bibitem{Belotsky:2018wph}
K.M.~Belotsky, V.I.~Dokuchaev, Y.N.~Eroshenko, E.A.~Esipova, M.Y.~Khlopov,
  L.A.~Khromykh et~al., \emph{{Clusters of primordial black holes}},
  \href{https://doi.org/10.1140/epjc/s10052-019-6741-4}{\emph{Eur. Phys. J. C}
  {\bfseries 79} (2019) 246}
  [\href{https://arxiv.org/abs/arXiv:1807.06590}{{\ttfamily
  arXiv:1807.06590}}].

\bibitem{Colazo:2024jmz}
P.E.~Colazo, F.~Stasyszyn and N.~Padilla, \emph{{Structure formation with
  primordial black holes to alleviate early star formation tension revealed by
  JWST}}, \href{https://doi.org/10.1051/0004-6361/202449565}{\emph{Astron.
  Astrophys.} {\bfseries 685} (2024) L8}
  [\href{https://arxiv.org/abs/arXiv:2404.13110}{{\ttfamily
  arXiv:2404.13110}}].

\bibitem{Khlopov:2004sc}
M.Y.~Khlopov, S.G.~Rubin and A.S.~Sakharov, \emph{{Primordial structure of
  massive black hole clusters}},
  \href{https://doi.org/10.1016/j.astropartphys.2004.12.002}{\emph{Astropart.
  Phys.} {\bfseries 23} (2005) 265}
  [\href{https://arxiv.org/abs/astro-ph/0401532}{{\ttfamily
  astro-ph/0401532}}].

\bibitem{Barcelo-Sastre:2026nbs}
M.~Barcel{\'o}-Sastre and J.~Garc{\'\i}a-Bellido, \emph{{Gravitational-wave
  signatures of primordial black hole clusters and the imprint of an early
  dense-core collapse}},
  \href{https://arxiv.org/abs/arXiv:2609.01331}{{\ttfamily arXiv:2609.01331}}.

\bibitem{Zhang:2026uum}
B.~Zhang, W.-X.~Feng and H.~An, \emph{{Small-Scale Clustering of Primordial
  Black Holes: The Little Red Dot Mass Function and the High-Redshift Galaxy
  Tension}},  \href{https://arxiv.org/abs/arXiv:2609.09078}{{\ttfamily
  arXiv:2609.09078}}.

\bibitem{Desjacques:2018wuu}
V.~Desjacques and A.~Riotto, \emph{{Spatial clustering of primordial black
  holes}}, \href{https://doi.org/10.1103/PhysRevD.98.123533}{\emph{Phys. Rev.
  D} {\bfseries 98} (2018) 123533}
  [\href{https://arxiv.org/abs/arXiv:1806.10414}{{\ttfamily
  arXiv:1806.10414}}].

\bibitem{Matsubara:2019qzv}
T.~Matsubara, T.~Terada, K.~Kohri and S.~Yokoyama, \emph{{Clustering of
  primordial black holes formed in a matter-dominated epoch}},
  \href{https://doi.org/10.1103/PhysRevD.100.123544}{\emph{Phys. Rev. D}
  {\bfseries 100} (2019) 123544}
  [\href{https://arxiv.org/abs/arXiV:1909.04053}{{\ttfamily
  arXiV:1909.04053}}].

\bibitem{Kawasaki:2021zir}
M.~Kawasaki, K.~Murai and H.~Nakatsuka, \emph{{Strong clustering of primordial
  black holes from Affleck-Dine mechanism}},
  \href{https://doi.org/10.1088/1475-7516/2021/10/025}{\emph{J. Cosmology
  Astropart. Phys.} {\bfseries 10} (2021) 025}
  [\href{https://arxiv.org/abs/arXiv:2107.03580}{{\ttfamily
  arXiv:2107.03580}}].

\bibitem{Mishra:2019pzq}
S.S.~Mishra and V.~Sahni, \emph{{Primordial Black Holes from a tiny bump/dip in
  the Inflaton potential}},
  \href{https://doi.org/10.1088/1475-7516/2020/04/007}{\emph{J. Cosmology
  Astropart. Phys.} {\bfseries 04} (2020) 007}
  [\href{https://arxiv.org/abs/arXiv:1911.00057}{{\ttfamily
  arXiv:1911.00057}}].

\bibitem{Saikawa:2018rcs}
K.~Saikawa and S.~Shirai, \emph{{Primordial gravitational waves, precisely: The
  role of thermodynamics in the Standard Model}},
  \href{https://doi.org/10.1088/1475-7516/2018/05/035}{\emph{J. Cosmology
  Astropart. Phys.} {\bfseries 05} (2018) 035}
  [\href{https://arxiv.org/abs/arXiv:1803.01038}{{\ttfamily
  arXiv:1803.01038}}].

\bibitem{Sasaki:2018dmp}
M.~Sasaki, T.~Suyama, T.~Tanaka and S.~Yokoyama, \emph{{Primordial black
  holes{\textemdash}perspectives in gravitational wave astronomy}},
  \href{https://doi.org/10.1088/1361-6382/aaa7b4}{\emph{Class. Quant. Grav.}
  {\bfseries 35} (2018) 063001}
  [\href{https://arxiv.org/abs/arXiv:1801.05235}{{\ttfamily
  arXiv:1801.05235}}].

\bibitem{Inman:2019wvr}
D.~Inman and Y.~Ali-Ha{\"\i}moud, \emph{{Early structure formation in
  primordial black hole cosmologies}},
  \href{https://doi.org/10.1103/PhysRevD.100.083528}{\emph{Phys. Rev. D}
  {\bfseries 100} (2019) 083528}
  [\href{https://arxiv.org/abs/arXiv:1907.08129}{{\ttfamily
  arXiv:1907.08129}}].

\bibitem{Lewis:1999bs}
A.~Lewis, A.~Challinor and A.~Lasenby, \emph{{Efficient computation of CMB
  anisotropies in closed FRW models}},
  \href{https://doi.org/10.1086/309179}{\emph{Astrophys. J.} {\bfseries 538}
  (2000) 473} [\href{https://arxiv.org/abs/astro-ph/9911177}{{\ttfamily
  astro-ph/9911177}}].

\bibitem{Iocco:2024rez}
F.~Iocco and L.~Visinelli, \emph{{Compatibility of JWST results with exotic
  halos}}, \href{https://doi.org/10.1016/j.dark.2024.101496}{\emph{Phys. Dark
  Univ.} {\bfseries 44} (2024) 101496}
  [\href{https://arxiv.org/abs/arXiv:2403.13068}{{\ttfamily
  arXiv:2403.13068}}].

\bibitem{Sheth:1999mn}
R.K.~Sheth and G.~Tormen, \emph{{Large scale bias and the peak background
  split}}, \href{https://doi.org/10.1046/j.1365-8711.1999.02692.x}{\emph{Mon.
  Not. Roy. Astron. Soc.} {\bfseries 308} (1999) 119}
  [\href{https://arxiv.org/abs/astro-ph/9901122}{{\ttfamily
  astro-ph/9901122}}].

\bibitem{Sheth:1999su}
R.K.~Sheth, H.J.~Mo and G.~Tormen, \emph{{Ellipsoidal collapse and an improved
  model for the number and spatial distribution of dark matter haloes}},
  \href{https://doi.org/10.1046/j.1365-8711.2001.04006.x}{\emph{Mon. Not. Roy.
  Astron. Soc.} {\bfseries 323} (2001) 1}
  [\href{https://arxiv.org/abs/astro-ph/9907024}{{\ttfamily
  astro-ph/9907024}}].

\bibitem{Press:1973iz}
W.H.~Press and P.~Schechter, \emph{{Formation of galaxies and clusters of
  galaxies by selfsimilar gravitational condensation}},
  \href{https://doi.org/10.1086/152650}{\emph{Astrophys. J.} {\bfseries 187}
  (1974) 425}.

\bibitem{Schneider:2020xmf}
A.~Schneider, S.K.~Giri and J.~Mirocha, \emph{{Halo model approach for the
  21-cm power spectrum at cosmic dawn}},
  \href{https://doi.org/10.1103/PhysRevD.103.083025}{\emph{Phys. Rev. D}
  {\bfseries 103} (2021) 083025}
  [\href{https://arxiv.org/abs/arXiv:2011.12308}{{\ttfamily
  arXiv:2011.12308}}].

\end{thebibliography}\endgroup
\bibliographystyle{JHEP}

\end{document}